\documentclass[a4paper,fleqn,usenatbib]{mnras}

\usepackage{mathptmx}

\usepackage[T1]{fontenc}
\usepackage{ae,aecompl}

\usepackage{graphicx}	
\usepackage{amsmath}	
\usepackage{amssymb}	
\usepackage{bigints}    
\usepackage{upgreek}
\usepackage{bm}
\usepackage{xspace} 
\usepackage{mn2e-breakabs}
\hypersetup{draft}

\def\halogen{{\sc{Halogen}}\xspace}
\def\cola{{\sc{ICE-COLA}}\xspace}
\def\patchy{{\sc{Patchy}}\xspace}
\def\pin{{\sc{Pinocchio}}\xspace}
\def\peak{{\sc{Peak Patch}}\xspace}

\newcommand*{\dint}{\, \mathrm{d}}

\title[Correlation Function Covariance comparison]{Comparing approximate methods for mock catalogues and covariance matrices I: correlation function}

\author[M. Lippich et al.]{Martha Lippich$^{1,2}$,\thanks{mlippich@mpe.mpg.de}
Ariel G. S\'{a}nchez$^{2}$,
Manuel Colavincenzo$^{3,4,5}$,
\newauthor
Emiliano Sefusatti$^{6,7}$,
Pierluigi Monaco$^{5,6,7}$,
Linda Blot$^{8,9}$,
Martin Crocce$^{8,9}$,
\newauthor
Marcelo A. Alvarez$^{10}$,
Aniket Agrawal$^{11}$,
Santiago Avila$^{12}$,
\newauthor
Andr\'es Balaguera-Antol\'{i}nez$^{13,14}$,
Richard Bond$^{15}$,
Sandrine Codis$^{15,16}$,
\newauthor
Claudio Dalla Vecchia$^{13,14}$,
Antonio Dorta$^{13,14}$,
Pablo Fosalba$^{8,9}$,
\newauthor
Albert Izard$^{17,18,8,9}$,
Francisco-Shu Kitaura$^{13,14}$,
Marcos Pellejero-Ibanez$^{13,14}$,
\newauthor
George Stein$^{15}$,
Mohammadjavad Vakili$^{19}$,
Gustavo Yepes$^{20,21}$
\\
\vspace{0.1cm}\\
$^{1}$ Universit\"ats-Sternwarte M\"unchen, Ludwig-Maximilians-Universit\"at M\"unchen, Scheinerstrasse 1, 81679 Munich, Germany\\
$^{2}$ Max-Planck-Institut f\"ur extraterrestrische Physik, Postfach 1312, Giessenbachstr., 85741 Garching, Germany\\
$^{3}$ Dipartimento  di  Fisica,  Universit\`a  di  Torino,  Via  P.  Giuria  1,  10125  Torino,  Italy\\
$^{4}$ Istituto  Nazionale  di  Fisica  Nucleare,  Sezione  di  Torino,  Via  P.  Giuria  1,  10125  Torino,  Italy\\
$^{5}$ Dipartimento di Fisica, Sezione di Astronomia, Universit\`a di Trieste, via Tiepolo 11, 34143 Trieste, Italy\\
$^{6}$ Istituto  Nazionale  di  Astrofisica, Osservatorio Astronomico di Trieste, via Tiepolo 11, 34143 Trieste, Italy\\
$^{7}$ Istituto  Nazionale  di  Fisica  Nucleare, Sezione di Trieste, Via Valerio, 2, I-34127 Trieste, Italy\\
$^{8}$ Institute of Space Sciences (ICE, CSIC), Campus UAB, Carrer de Can Magrans, s/n,  08193 Barcelona, Spain \\
$^{9}$ Institut d'Estudis Espacials de Catalunya (IEEC), 08193 Barcelona, Spain \\
$^{10}$ Berkeley Center for Cosmological Physics, Campbell Hall 341, University of California, Berkeley CA 94720 \\
$^{11}$ Max-Planck-Institut f\"ur Astrophysik, Karl-Schwarzschild-Str. 1, 85741 Garching, Germany\\
$^{12}$ Institute of Cosmology \& Gravitation, Dennis Sciama Building, University of Portsmouth, Portsmouth PO1 3FX, UK\\
$^{13}$ Instituto de Astrof\'isica de Canarias, C/V\'ia  L\'actea, s/n, E-38200, La Laguna, Tenerife, Spain\\
$^{14}$ Departamento Astrof\'isica, Universidad de La Laguna,  E-38206 La Laguna, Tenerife, Spain\\
$^{15}$ Canadian Institute for Theoretical Astrophysics, University of Toronto, 60 St. George Street, Toronto, ON M5S 3H8, Canada\\
$^{16}$  Institut d'Astrophysique de Paris, CNRS \& Sorbonne Universit\'e, UMR 7095, 98 bis boulevard Arago, 75014 Paris, France\\
$^{17}$ Jet Propulsion Laboratory, California Institute of Technology, 4800 Oak Grove Drive, Pasadena, CA 91109, USA\\
$^{18}$ Department of Physics and Astronomy, University of California, Riverside, CA 92521, USA \\
$^{19}$ Leiden Observatory, Leiden University, P.O. Box 9513, NL-2300 RA, Leiden, The Netherlands \\
$^{20}$ Departamento de F\'isica Te\'orica, M\'odulo 15, Universidad Aut\'onoma de Madrid, 28049 Madrid, Spain\\
$^{21}$ Centro de Investigaci\'on  Avanzada en F\'{i}sica Fundamental (CIAFF), Universidad Aut\'onoma de Madrid, 28049, Madrid, Spain
}

\date{Accepted XXX. Received YYY; in original form ZZZ}

\pubyear{2018}

\begin{document}
\label{firstpage}
\pagerange{\pageref{firstpage}--\pageref{lastpage}}
\maketitle

\begin{abstract}
This paper is the first in a set that analyses the covariance matrices of clustering statistics obtained from several approximate methods for gravitational structure formation. We focus here on the covariance matrices 
of anisotropic two-point correlation function measurements.
Our comparison includes seven approximate methods, which can be divided into three 
categories: predictive methods that follow the evolution of the linear density 
field deterministically (\cola, \peak, 
and \pin), methods that require a calibration with N-body simulations 
(\patchy and \halogen), and simpler recipes based on assumptions regarding the shape 
of the probability distribution function (PDF) of density fluctuations 
(log-normal and Gaussian density fields).
We analyse the impact of using covariance estimates obtained from these approximate
methods on cosmological analyses of galaxy clustering measurements,
using as a reference the covariances inferred from a set of full N-body simulations.
We find that all approximate methods can accurately recover the mean 
parameter values inferred using the N-body covariances. The obtained parameter 
uncertainties typically agree with the corresponding N-body results within 5\% for our lower mass threshold, and 10\% for our higher mass threshold.
Furthermore, we find that the constraints for some methods
can differ by up to 20\% depending on whether the halo samples used to define the 
covariance matrices are defined by matching the mass, number density, or clustering
amplitude of the parent N-body samples.
The results of our configuration-space analysis indicate that most approximate 
methods provide similar results, with no single method clearly outperforming the 
others. 
\end{abstract}




\section{Introduction}

The statistical analysis of the large-scale structure (LSS) of the 
Universe is one of the primary tools of observational cosmology.
The analysis of the signature of baryon acoustic oscillations (BAO) and 
redshift-space distortions (RSD) on anisotropic two-point clustering measurements
can be used to infer constraints on the expansion history of the Universe 
\citep{Blake2003,Linder2003} and the redshift evolution of the growth-rate of 
cosmic structures \citep{Guzzo2008}. 
Thanks to this information, LSS observations have shaped our current understanding 
of some of the most challenging open problems in cosmology, such as the 
nature of dark energy, the behaviour of gravity on large scales, and the physics of
inflation \citep[e.g.][]{Efstathiou2002,Eisenstein2005, Cole2005, Sanchez2006,
Sanchez2012, Anderson2012, Anderson2013, Anderson2014, Alam2017}. 

Future galaxy surveys such as {\it Euclid} \citep{Laureijs2011} or the Dark Energy 
Spectroscopic Instrument (DESI) Survey \citep{desipaper} will contain millions of
galaxies covering large cosmological volumes. The small statistical uncertainties 
associated with clustering measurements based on these samples will push the precision 
of our tests of the standard $\Lambda$CDM scenario even further. 
In this context, it is essential to identify all components of the systematic error 
budget affecting cosmological analyses based on these measurements, as well as to 
define strategies to control or mitigate them. 

A key ingredient to extract cosmological information out of anisotropic clustering 
statistics is robust estimates of their covariance matrices. 
In most analyses, covariance matrices are computed from a set of mock catalogues 
designed to reproduce the properties of a given survey. 
Ideally, these mock catalogues should be based on N-body simulations, which can reproduce 
the impact of non-linear structure formation on the clustering properties of
a sample with high accuracy. 
Due to the finite number of mock catalogues, the estimation of the covariance matrix 
is affected 
by statistical errors and the resulting noise must be propagated into the final
cosmological constraints \citep{Taylor2013, Dodelson2013, Percival2014, Sellentin2016}.
Reaching the level of statistical precision needed for future surveys 
might require the generation of several thousands of mock catalogues. 
As N-body simulations are expensive in terms of run-time and memory, the construction
of a large number of mock catalogues might be infeasible. 
The required number of realizations can be reduced by means of  
methods such as resampling the phases of N-body simulations 
\citep{Hamilton2006, Schneider2011}, 
shrinkage \citep{Pope2008}, calibrating the non-Gaussian contributions of an empirical model against N-body simulations \citep{Connell2016},
or covariance tapering \citep{Paz2015}. 
However, even after applying such methods, the generation of multiple N-body 
simulations with the required number-density and volume to be used for 
the clustering analysis of future surveys 
would be extremely demanding.

During the last decades, several approximate methods for gravitational structure formation 
and evolution have been developed, which allow for a faster generation of mock catalogues; 
see \cite{Monaco2016} for a review.
The accuracy with which these methods reproduce the covariance matrices estimated from
N-body simulations must be thoroughly tested to avoid introducing systematic errors 
or biases on the parameter constraints derived from LSS measurements. 

The nIFTy comparison project by \citet{Chuang2015} presented a detailed comparison of 
major approximate methods regarding their ability to reproduce clustering statistics 
(two-point correlation function, power spectrum and bispectrum) of halo samples drawn
out of N-body simulations.
Here we take the comparison of different approximate methods one step further. We 
compare the covariance matrices inferred from halo samples obtained from different 
approximate methods to the corresponding ones derived from full N-body simulations. 
Furthermore, we also test the performance of the different covariance matrices at 
reproducing parameter constraints obtained using N-body simulations. 
We include seven approximate methods, which can be divided into three classes: predictive 
methods that evolve the linear density field deterministically on Lagrangian trajectories, 
including \cola \citep{Tassev2013, Izard2016}, \peak \citep{1996ApJS..103....1B}, and 
\pin \citep{Monaco2002,Munari2017}, methods that require higher calibration with 
N-body simulations, such as \halogen \citep{Avila2015}, and \patchy \citep{Kitaura2014}, 
and two simpler recipes based on models of the PDF of the density fluctuations, 
the Gaussian recipes of \citet{Grieb2016}, and realizations of log-normal density 
fields constructed using the code of \citep{Agrawal2017}. 
For the predictive and calibrated methods, we generate the same number of halo catalogues 
as the reference N-body simulations using identical initial conditions. 
We focus here on the comparison of the covariance matrices of two-point anisotropic 
clustering measurements in configuration space, considering 
Legendre multipoles \citep{Padmanabhan2008} and clustering wedges \citep*{Kazin2012}.
Our companion papers \citet{Blot18} and \cite{Colavincenzo18} perform an analogous 
comparison based on power spectrum and bispectrum measurements.

The structure of the paper is as follows. 
Section~\ref{sec:methods} presents a brief description of the reference N-body 
simulations and the different approximate methods and recipes included in our comparison. 
In Section~\ref{sec:methodology} we summarize the methodology used in this analysis, 
including a description of the halo samples that we consider (Section~\ref{sec:samples}), 
our clustering measurements (Section~\ref{sec:2pcf}), the estimation of the corresponding 
covariance matrices (Section~\ref{sec:cov-mat-est}),
and the modelling for the correlation function used to asses the impact of the 
different methods when estimating parameter constraints (Section~\ref{sec:xi_model}). 
We present a comparison of the clustering properties of the different 
halo samples in Section~\ref{sec:cf-comp} and their corresponding covariance matrices
in Section~\ref{sec:cov_mat_comp}. 
In Section~\ref{sec:performance-comp} we compare the performance of the different covariance 
matrices by analysing parameter constraints obtained from representative fits, using 
as a reference the ones obtained when the analysis is based on N-body simulations. 
We discuss the results from this comparison in Section~\ref{sec:discussion}.
Finally, Section~\ref{sec:conclusions} presents our main conclusions.

\section{Approximate methods for covariance matrix estimates}
\label{sec:methods}

\subsection{Methods included in the comparison}

In this comparison project, we included covariance matrices inferred from 
different approximate methods and recipes, which we 
compared to the estimates obtained from a set of reference N-body simulations. 
Approximate methods have recently been revived by high-precision cosmology, due to the 
need of producing a large number of realizations 
to compute covariance matrices of clustering measurements. 
This topic has been reviewed by \cite{Monaco2016}, where methods have been roughly divided 
into two broad classes. 
``Lagrangian" methods, as N-body simulations, are applied to a grid of particles 
subject to a perturbation field. They reconstruct the Lagrangian patches that collapse 
into dark matter halos, and then displace them to their Eulerian positions at the output 
redshift, typically with Lagrangian Perturbation Theory (hereafter LPT). \cola, 
\peak and \pin fall in this class. These methods are predictive, in 
the sense that, after some cosmology-independent calibration of their free parameters 
(that can be thought at the same level as the linking length of friends-of-friends halo 
finders), they give their best reproduction of halo masses and clustering without any 
further tuning. This approach can be demanding in terms of computing resources and can have high 
memory requirements. In particular, \cola belongs to the class of Particle-Mesh codes; 
these are in fact N-body codes that converge to the true solution (at least on 
large scales) for sufficiently small time-steps. As such, Particle-Mesh codes are expected to 
be more accurate than other approximate methods, at the expense of higher computational costs.

The second class of ``bias-based'' methods is based on the idea of creating a mildly 
non-linear density field using some version of LPT, and then populate the density field with 
halos that follow a given mass function and a specified bias model. The 
parameters of the bias model must be calibrated on a simulation, so as to reproduce halo 
clustering as accurately as possible. The point of strength of these methods is their very 
low computational cost and memory requirement, that makes it possible to generate thousands 
of realizations in a simple workstation, and to push the mass limit to very low masses. 
This is however achieved at the cost of lower predictivity, and need of recalibration 
when the sample selection changes. \halogen and \patchy fall in this category.

In the following, we will refer to the two classes as ``predictive'' and ``calibrated'' models. 
All approximate methods used here have been applied to the same set of 300 initial 
conditions (ICs) of the reference N-body simulations, so as to be subject to the same 
sample variance; 
as a consequence, the comparison, though limited to a relatively small number of 
realizations, is not affected by sample variance. 

Additionally, we included in the comparison two simple recipes for the shape of the 
PDF of the density fluctuations, a Gaussian analytic 
model that is only valid in linear theory and a log-normal model. The latter was implemented 
by generating 1000 catalogues of ``halos'' that Poisson-sample a log-normal density field; 
in this test case we do not match the ICs with the reference simulations, and use a 
higher number of realizations to lower sample variance.

\subsection{Reference N-body halo catalogue: Minerva}
\label{sec:minerva}

Our reference catalogues for the comparison of the different approximate methods is 
derived from a set of 300 N-body simulations called Minerva, which were performed 
using GADGET-3 \citep[last described in][]{Springel2005}.
To the first set of 100 realizations, which is described in more detail in \citet{Grieb2016} 
and was used in the recent BOSS analyses by \citet{Sanchez2017} and \citet{Grieb2017}, 
200 new independent realizations were added, which were generated 
with the same set-up as the first simulations.
The initial conditions were derived from second-order Lagrangian perturbation theory (2LPT) 
and use the cosmological parameters that match the best-fitting results of the 
WMAP+BOSS DR9 analysis by \citet{Sanchez2013} at a starting redshift $z_{\text{ini}}=63$.
Each realization is a cubic box of side length $L_{\text{box}} =1.5\,h^{-1}$Gpc with 
$1000^3$ dark-matter (DM) particles and periodic boundary conditions.
For the approximate methods described in the following sections we use the same box size 
and exactly the same ICs for each realization as in the Minerva simulations.
Halos were identified with a standard Friends-of-friends (FoF)
algorithm at a snapshot of the simulations at $z=1.0$. FoF halos were then subject to the unbinding procedure
provided by the {\sc SUBFIND} code \citep{Springel2001}, where particles with positive
total energy are removed and halos that were artificially linked by
FoF are separated.
Given the particle mass resolution of the Minerva simulations, the minimum halo mass is $2.667\times 10^{12}$\,$h^{-1}$M$_{\sun}$.

\subsection{Predictive methods}
\subsubsection{\cola}
\label{sec:cola}

{\sc COLA} \citep{Tassev2013} is a method to speed up N-body simulations by incorporating 
a theoretical modelling of the dynamics into the N-body solver and using a low 
resolution numerical integration. It starts by computing the initial conditions using 
second-order Lagrangian Perturbation Theory (2LPT, see \citealt{aCrocce2006}). Then, it 
evolves particles along their 2LPT trajectories and adds a residual displacement with 
respect to the 2LPT path, which is integrated numerically using the N-body solver. 
Mathematically, the displacement field $\bmath{x}$ is decomposed into the LPT component 
$\bmath{x}_{LPT}$ and the residual displacement $\bmath{x}_{\rm res}$ as
\begin{equation}
  \bmath{x}_{\rmn{res}}(t) \equiv \bmath{x}(t)-\bmath{x}_{\rmn{LPT}}(t).
  \label{eq:cola_displacement_field}
\end{equation}
In a dark matter-only simulation, the equation of motion relates the acceleration to 
the Newtonian potential $\Phi$, and omitting some constants it can be written as: 
$\partial_{t}^{2} \bmath{x}(t) = -\nabla\Phi(t)$. Using 
equation~(\ref{eq:cola_displacement_field}), the equation of motion reads
\begin{equation}
  \partial_{t}^{2}\bmath{x}_{\rmn{res}}(t) = -\nabla\Phi(t)-\partial_{t}^{2}\bmath{x}_{\rmn{LPT}}(t).
\end{equation}
{\sc COLA} uses a Particle-Mesh method to compute the gradient of the potential at the position 
$\bmath{x}$ (first term of the right hand side), it subtracts the acceleration corresponding 
to the LPT trajectory and finally the time derivatives on the left hand side are discretized 
and integrated numerically using few time steps. The 2LPT ensures convergence of the dynamics 
at large scales, where its solution is exact, and the numerical integration solves the 
dynamics at small non-linear scales. Halos can be correctly identified running a 
Friends-of-Friends (FoF) algorithm \citep{Davis1985} on the dark matter density field, 
and halo masses, positions and velocities are recovered with accuracy enough to build 
mock halo catalogues.

\cola \citep{Izard2016,Izard2018} is a modification of the parallel version of 
{\sc COLA} developed in \citet{Koda2016} that produces all-sky light cone catalogues on-the-fly. 
\cite{Izard2016} presented an optimal configuration for the production of accurate mock 
halo catalogues and \citet{Izard2018} explains the light cone production and the 
modelling of weak lensing observables.

Mock halo catalogues were produced with \cola placing 30 time steps between an initial 
redshift of $z_i=19$ and $z=0$\footnote{The time steps were linearly distributed with the 
scale factor.} and forces were computed in a grid with a cell size 3 times smaller than the 
mean inter-particle separation distance. For the FoF algorithm, a linking length of $b=0.2$ 
was used. Each simulation reached redshift 0 and used 200 cores for 20 minutes in 
the MareNostrum3 supercomputer at the Barcelona Supercomputing Center
\footnote{\url{http:www.bsc.es}.}, consuming a total of 20 CPU khrs for the 300 
realizations.

\subsubsection{\peak}
\label{sec:peakpatch}

From each of the 300 initial density field maps of the Minerva suite, we generate 
halo catalogues following the peak patch approach initially introduced by 
\citet{1996ApJS..103....1B}. In particular, we use a new massively parallel implementation 
of the peak patch algorithm 
to create efficient and accurate realizations of the positions and peculiar velocities of dark 
matter halos (Stein, Alvarez, and Bond 2018, in prep.). The peak patch approach is essentially a Lagrangian space halo finder 
that associates halos with the largest regions that have just collapsed by a given time. 
The pipeline can be separated into four subprocesses: (1) the generation of a random 
linear density field with the same phases and power spectrum as the Minerva simulations; 
(2) identification of collapsed regions using the homogeneous ellipsoidal collapse 
approximation; (3) exclusion and merging of the collapsed regions in Lagrangian space; 
and (4) assignment of displacements to these halos using second order Lagrangian 
perturbation theory.

The identification of collapsed regions is a key step of the algorithm. The determination 
of whether any given region will have collapsed or not is made by approximating it as 
an homogeneous ellipsoid, the fate of which is determined completely by the principal axes 
of the deformation tensor of the linear displacement field (i.e. the strain) averaged over 
the region. In principle, the process of finding these local mass peaks would involve 
measuring the strain at every point in space, smoothed on every scale. However, 
experimentation has shown that equivalent results can be obtained by measuring the strain 
around density peaks found on a range of scales\footnote{This is not to say that a halo 
found on a given scale corresponds to a peak in the density smoothed on that scale, 
however, which is only the case when the strain is isotropic and the collapse is spherical. 
Thus, the use of density peaks as centers for strain measurements and ellipsoidal 
collapse calculations in the algorithm is only an optimization, to avoid wasting 
computations measuring the properties of regions of Lagrangian space that are unlikely to 
collapse in the first place. }. This is done by smoothing the field on a series of 
logarithmically spaced scales with a top-hat kernel, from a minimum radius of 
$R_{f,{\rm min}} = 2a_{\rm latt}$, where $a_{\rm latt}$ is the lattice spacing, to a 
maximum radius of $R_{f,{\rm max}} = 40\,{\rm Mpc}$, with a ratio of 1.2.  
For each candidate peak, we then find the largest radius for which a homogeneous 
ellipsoid with the measured mean strain would collapse by the redshift of interest. 
If a candidate peak has no radius for which a homogeneous ellipsoid with the measured 
strain would have collapsed, then that point is thrown out. Each candidate point is 
then stored as a peak patch at its location with its radius. We then proceed down through 
the filter bank to all scales and repeat this procedure for each scale, resulting in a list 
of peak patches which we refer to as the unmerged catalogue.

The next step is to  account for exclusion, an essential step to avoid double counting of 
matter, since distinct halos should not overlap, by definition. We choose here to use 
binary exclusion \citep{1996ApJS..103....1B}. Binary exclusion starts from a ranked list 
of candidate peak patches sorted by mass or, equivalently, Lagrangian peak patch radius. 
For each patch we consider every other less massive patch that overlaps it. If the 
smaller patch is outside of the larger one, then the radius of the two patches is reduced 
until they are just touching. If the center of the smaller patch is inside the large one, 
then that patch is removed from the list. This process is repeated until the least massive 
remaining patch is reached.

Finally, we move halos according to 2LPT using displacements computed at the scale of 
the halo.

This method is very fast: each realization ran typically in 97 seconds on 64 cores
of the GPC supercomputer at the SciNet HPC Consortium in Toronto (1.72 hours in total) 
. It allows to get accurate -- and fast -- halo 
catalogues without any calibration, achieving high precision on the mass function 
typically for masses above a few $10^{13} M_{\odot}$. 

\subsubsection{\pin}
\label{sec:pino}

The PINpointing Orbit Crossing Collapsed HIerarchical Objects (\pin) code 
\citep{Monaco2002} is based on the following algorithm.

A linear density contrast field is generated in Fourier space, in a way similar to 
N-body simulations. As a matter of fact, the code version used here implements the same 
loop in $k$-space as the initial condition generator (N-GenIC) used for the simulations, 
so the same realization is produced just by providing the code with the same random seed. 
The density is then smoothed using several smoothing radii. For each smoothing radius, 
the code computes the time at which each grid point (``particle'') is expected to get to 
the highly non-linear regime. The dynamics of grid points, as mass elements, is treated 
as the collapse of a homogeneous ellipsoid, whose tidal tensor is given by the Hessian 
of the potential at that point. Collapse is defined as the time at which the ellipsoid 
collapses on the first axis, going through orbit crossing and into the highly non-linear 
regime; this is a difference with respect to \peak, where the collapse of extended 
structures is modelled. The equations for ellipsoidal collapse are solved using third-order 
Lagrangian Perturbation Theory (3LPT). Following the ideas behind excursion-sets theory, 
for each particle we consider the earliest collapse time as obtained by varying the 
smoothing radius.

Collapsed particles are then grouped together using an algorithm that mimics the 
hierarchical assembly of halos: particles are addressed in chronological order of 
collapse time; when a particle collapses the six nearest neighbours in the Lagrangian 
space are checked, if none has collapsed yet then the particle is a peak of the 
inverse collapse time (defined as $F=1/D_c$, where $D_c = D(t_c)$ is the growth rate at 
the collapse time) and it becomes a new halo of one particle. If the collapsed particle 
is touching (in the Lagrangian space) a halo, then both the particle and the halo 
are displaced using LPT, and if they get ``near enough'' the particle is accreted to the 
halo, otherwise it is considered as a ``filament'' particle, belonging to the filamentary 
network of particles that have suffered orbit crossing but do not belong to halos. If a 
particle touches two halos, then their merging is decided by moving them and checking 
whether they get again ``near enough''. Here ``near enough'' implies a parametrization that 
is well explained in the original papers 
\citep[see][for the latest calibration]{Munari2017}. This results in the construction of 
halos together with their merger histories, obtained with continuous time sampling. 
Halos are then moved to the final position using 3LPT.
The so-produced halos have discrete masses, proportional to the
particle mass $M_p$, as the halos found in N-body simulations. To ease
the procedure of number density matching described below in Section 3,
halo masses were made continuous using the following procedure. It is
assumed that a halo of $N$ particles has a mass that is distributed
between $N \times M_p$ and $(N+1) \times M_p$, and the distribution is
obtained by interpolating the mass function as a power law between two
values computed in successive bins of width $M_p$. This procedure
guarantees that the cumulative mass function of halos of mass $>
N\times M_p$ does not change, but it does affect the differential mass
function.

We use the latest code version presented in \cite{Munari2017}, where the advantage of 
using 3LPT is demonstrated. No further calibration was required before starting the 
runs. That paper presents scaling tests of the massively parallel version V4.1 and 
timings. The 300 runs were produced in the GALILEO@CINECA Tier-1 facility, each run 
required about 8 minutes on 48 cores.

\subsection{Calibrated methods}
\subsubsection{\halogen}
\label{sec:halogen}
\halogen \citep{Avila2015} is an approximate method designed to generate halo 
catalogues with the correct two-point correlation function as a function of mass. 
It constructs the catalogues following four simple steps:
\begin{itemize}
\item Generate a 2LPT dark matter field, and distribute their particles on a grid with 
cell size $l_{\rm cell} $.
\item Draw halo masses $M_h$ from an input Halo Mass Function (HMF). 
\item Place the halo masses (from top to bottom) in the cells with a probability that depends on the cell density and the halo mass $P\propto \rho_{\rm cell}^{\ \ \alpha(M_h)}$. Within cells we choose random particles to assign the halo position. We further ensure mass conservation within cells and avoid halo overlap.  
\item Assign halo velocities from the particle velocities, with a velocity bias factor:  ${\bf v}_{\rm halo} = f_{\rm vel} (M_h) \cdot {\bf v}_{\rm part}$
\end{itemize}

Following the study in \citep{Avila2015}, we fix the cell size at 
$l_{\rm cell}=5\,h^{-1}{\rm Mpc}$. In this paper we take the input HMF from the mean of the 
300 Minerva simulations, but in other studies analytical HMF have been used. The parameter 
$\alpha(M_h)$ controls the clustering as a function of halo mass and has been calibrated 
using the two-point function from the Minerva simulations in logarithmic mass bins ($M_{h}=1.06\times10^{13}$,  $2.0\times 10^{13}$, $4.0\times 10^{13}$, $8.0\times 10^{13}$, $1.6\times 10^{14}\,h^{-1}M_{\sun}$). 
The factor $f(M_h)$ is also tuned to match the variance of the halo velocities from the 
N-body simulations.

\halogen is a code that advocates for the simplicity and low needs of computing 
resources. The fact that it does not resolve halos (i.e. using a halo finder), allows to 
probe low halo masses while keeping low the computing resources. This has the disadvantage 
of needing to introduce free parameters. However, \halogen only needs one 
clustering parameter $\alpha$ and one velocity parameter $f_{\rm vel}$, making the 
fitting procedure simple.  

\subsubsection{\patchy}
\label{sec:patchy}
The \patchy code \citep{Kitaura2014,Kitaura:2014mja} relies on modelling the 
large-scale density field with an efficient approximate gravity solver, which is 
populated with the halo density field using a non-linear, scale dependent, and 
stochastic biasing description. Although it can be applied to directly paint the 
galaxy distribution on the density mesh \cite[see][]{Kitaura2016}.

The gravity solver used in this work is based on Augmented Lagrangian Perturbation 
Theory  \cite[ALPT,][]{Kitaura:2012tj}, fed with the same initial conditions as 
those implemented in the Minerva simulations. In the ALPT model, 2LPT is modified by 
employing a spherical collapse model on small comoving scales, splitting the 
displacement field into a long and a short range component. Better results can in 
principle be obtained using a particle mesh gravity solver at a higher computational 
cost \cite[see][]{Vakili:2017rsp}. 

Once the dark matter density field is computed, a deterministic bias relating it to 
the expected number density of halos is applied. This deterministic bias model consists 
of a threshold, an exponential cut-off, and a power-law bias relation. The number density 
is fixed by construction using the appropriate normalization of the bias expression.

The \patchy code then associates the number of halos in each cell by sampling from a 
negative binomial distribution modelling the deviation from Poissonity with an 
additional stochastic bias parameter.

In order to provide peculiar velocities, these are split into a coherent and 
a quasi-virialised component. The coherent flow is obtained from ALPT and the 
dispersion term is sampled from a Gaussian distribution assuming a power law with the 
local density.

The masses are associated to the halos by means of the {\sc HADRON} code \citep{Zhao:2015jga}. 
In this approach, the masses coming from the N-body simulation are classified in 
different density bins and in different cosmic web types (knots, filaments, sheets and 
voids) and their distribution information is extracted. Then {\sc HADRON} uses this information 
to assign masses to halos belonging to mock catalogues. This information is independent 
of initial conditions, meaning it will be the same for each of the 300 Minerva 
realizations.

We used the MCMC python wrapper published by \cite{Vakili:2017rsp} to infer the values of 
the bias parameters from Minerva simulations using one of the 300 random realizations. 
Once these parameters are fixed one can produce all of the other mock catalogues without 
further fitting. 
The \patchy mocks were produced using a down-sampled white noise of $500^3$ instead of 
the $1000^3$ original Minerva ones with an effective cell side resolution of 
$3\,h^{-1}{\rm  Mpc }$ to produce the dark matter field.

\subsection{Models of the density PDF}

\subsubsection{Log-normal distribution}
\label{sec:lognormal}

The log-normal mocks were produced using the public code presented 
in \citet{Agrawal2017}, which models the matter and halo density fields as log-normal 
fields, and generates the velocity field from the matter density field, using the 
linear continuity equation. 

To generate a log-normal field $\delta(\bm x)$, a Gaussian field $G(\bm x)$ is first 
generated, which is related to the log-normal field as 
$\delta(\bm x) = e^{-\sigma^2_G+G(\bm x)}-1$~\citep{Coles1991}. The pre-factor with 
the variance $\sigma^2_G$ of the Gaussian field $G(\bm x)$, ensures that the mean of 
$\delta(\bm x)$ vanishes. Because different Fourier modes of a Gaussian field 
are uncorrelated, the Gaussian field $G(\bm x)$ is generated in Fourier space. The 
power spectrum of $G(\bm x)$ is found by Fourier transforming its correlation function 
$\xi^G(r)$, which is related to the correlation function $\xi(r)$ of the log-normal field 
$\delta(\bm x)$ as $\xi^G(r) = \text{ln}[1+\xi(r)]$~\citep{Coles1991}. Having generated 
the Gaussian field $G(\bm x)$, the code transforms it to the log-normal field 
$\delta(\bm x)$ using the variance $\sigma^2_G$ measured from $G(\bm x)$ in all cells. 

In practice, we use the measured real-space matter power spectrum from Minerva and 
Fourier transform it to get the matter correlation function. For halos we use the 
measured real-space correlation function. We then generate the Gaussian matter and halo 
fields with the same phases, so that the Gaussian fields are perfectly correlated with 
each other. Note however, that we use random realizations for these mocks, and so, 
these phases are not equal to those of the Minerva initial conditions. We then 
exponentiate the Gaussian fields, to get matter ($\delta_m(\bm x)$) and halo 
($\delta_g(\bm x)$) density fields, following a log-normal distribution. 

The expected number of halos in a cell is given as 
$N_g(\bm x) = \bar{n}_g[1+\delta_g(\bm x)]V_{\text{cell}}$, where $\bar{n}_g$ is the 
mean number density of the halo sample from Minerva, $\delta_g(\bm x)$ is the halo 
density at position $\bm x$, and $V_{\text{cell}}$ is the volume of the cell. However, 
this is not an integer. So, to obtain an integer number of halos from the halo density 
field, we draw a random number from a Poisson distribution with mean $N_g(\bm x)$, and 
populate halos randomly within the cell. The log-normal matter field is then used to 
generate the velocity field using the linear continuity equation. Each halo in a cell 
is assigned the three-dimensional velocity of that cell. 

Since the log-normal mocks use random phases, we generate 1000 realizations for each mass 
bin, with the real-space clustering and mean number density measured from Minerva as 
inputs. Also note, that because halos in this prescription correspond to just discrete 
points, we do not assign any mass to them. An effective bias relation can still be 
established using the cross-correlation between the halo and matter fields, or using 
the input clustering statistics~(\cite{Agrawal2017}). 

The key advantage of using this method is its speed. Once we had the target power 
spectrum of the matter and halo Gaussian fields, each realization of a $256^3$ grid 
as in Minerva, was produced in $20$ seconds using $16$ cores at the RZG in Garching. 
The resulting catalogues agree perfectly with the Minerva realizations in their 
real-space clustering as expected. Because we use linear velocities, they also agree 
with the redshift-space predictions on large scales \citep{Agrawal2017}.  

\subsubsection{Gaussian distribution}
\label{sec:gaussian}

A different approach to generating ``mock'' halo catalogues with fast approximate methods 
is to model the covariance matrix theoretically.
This has the advantage that the resulting estimate is free of noise. 
In this comparison project we included a simple theoretical model for the linear 
covariance of anisotropic galaxy clustering that is described in \citet{Grieb2016}.
Based on the assumption that the two-dimensional power spectrum $P(k,\mu)$ follows a 
Gaussian distribution and that the contributions from the trispectrum and super-sample 
covariance can be neglected, \citet{Grieb2016} derived the explicit formulae for the 
covariance of anisotropic clustering measurements in configuration and Fourier space.
In particular, they obtain that the covariance between two Legendre multipoles of
the correlation function of order $\ell$ and $\ell'$ (see Section~\ref{sec:2pcf}) 
evaluated at the pair separations $s_i$ and $s_j$, respectively, is given by
\begin{equation}
 C_{\ell,\ell'}(s_i, s_j) = \frac{i^{\ell+\ell'}}{2\pi ^2}\int_0 ^{\infty} k^2 \sigma_{\ell\ell'}^2(k)\bar{j}_{\ell}(ks_i)\bar{j}_{\ell'}(ks_j)\dint k,
\end{equation}
where $\bar{j}_{\ell}(ks_i)$ is the bin-averaged spherical Bessel function as defined in 
equation A19 of \citet{Grieb2016}, and 
\begin{multline}
 \label{eq:ps_cov_ell_ell}
 \sigma^2_{\ell\ell'}(k) \equiv \frac{(2 \ell + 1) \, (2 \ell' + 1)}{V_\mathrm{s}} \\
 \times \int_{-1}^1 \left[ P(k, \mu) + \frac{1}{\bar n} \right]^2 L_{\ell}(\mu) \, L_{\ell'}(\mu) \dint \mu.
\end{multline}
Here, $P(k, \mu)$ represents the two-dimensional power spectrum of the sample, $V_\mathrm{s}$
is its volume, and $\bar{n}$ corresponds to its mean number density. 

 Analogously, the covariance between two configuration-space 
clustering wedges $\mu$ and $\mu'$ (see Section~\ref{sec:2pcf}) is given by 
\begin{equation}
\begin{split}
 C_{\mu,\mu'}(s_i, s_j) =  &\quad \sum _{\ell _1 \ell _2} \frac{i^{\ell_1+\ell_2}}{2\pi ^2}\bar{L} _{\ell_1, \mu } \bar{L} _{\ell_2, \mu '} \\
 &\quad \times \int_0 ^{\infty} k^2 \sigma_{\ell _1\ell _2}^2(k)j _{\ell_1}(ks_i)j _{\ell_2}(ks_j)\dint k,
 \end{split}
\end{equation}
where $\bar{L}_{\ell_1, \mu}$ represents the average of the Legendre polynomial 
of order $\ell$ within the corresponding $\mu$-range of the clustering wedge.
The covariance matrices derived from the Gaussian model have been tested against 
N-body simulations with periodic boundary conditions by \citet{Grieb2016}, 
showing good agreement within the range of scales typically included in 
the analysis of galaxy redshift surveys ($s > 20\,h^{-1}{\rm Mpc}$).

\section{Methodology}
\label{sec:methodology}

\subsection{Halo samples}
\label{sec:samples}

In this section we describe the criteria used to construct the halo samples on which 
we base our covariance matrix comparison.

\begin{figure*}
\begin{minipage}[c]{\columnwidth}
 \centering
 \includegraphics[width=0.9\columnwidth]{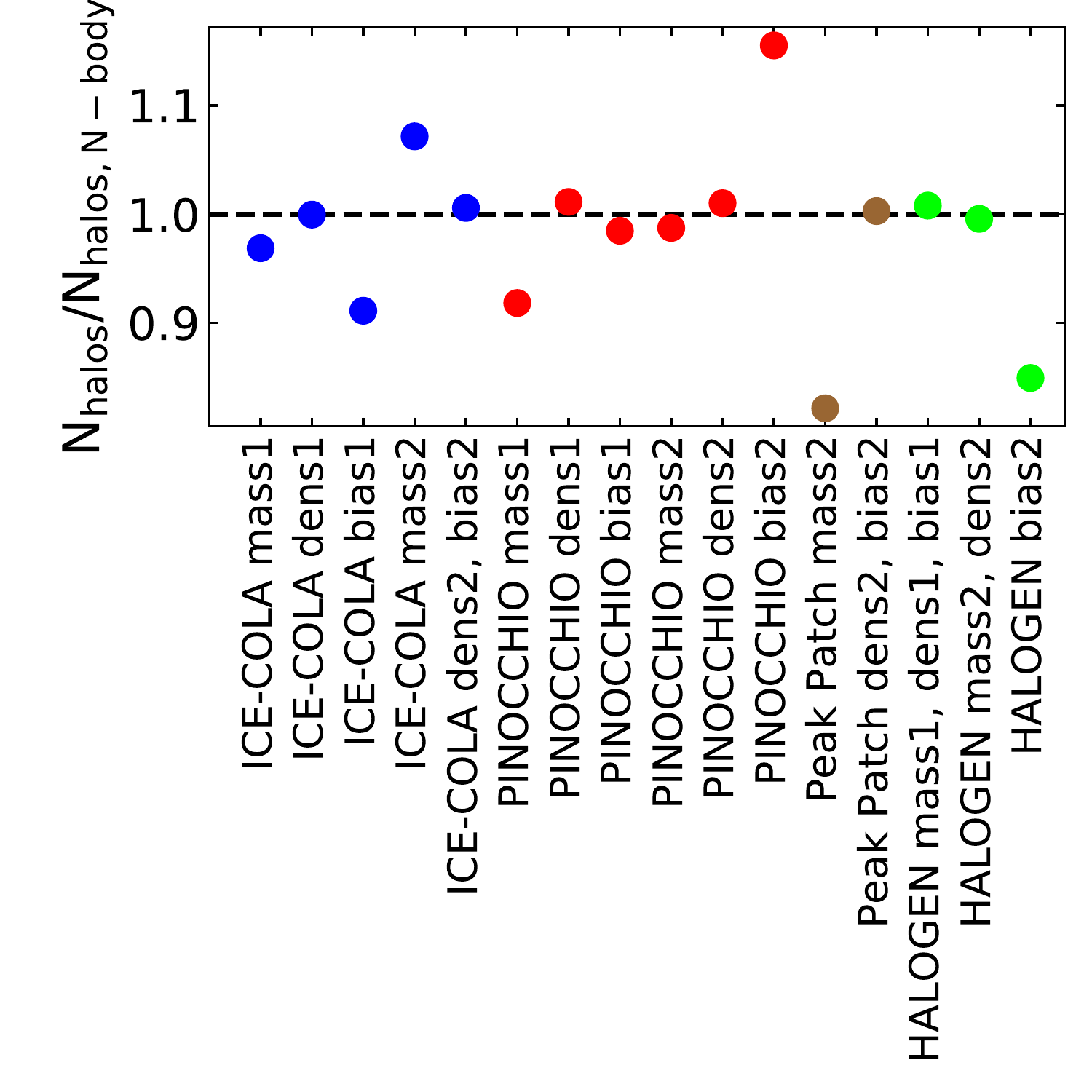}
 \end{minipage}
 \begin{minipage}[c]{\columnwidth}
  \centering
 \includegraphics[width=0.9\columnwidth]{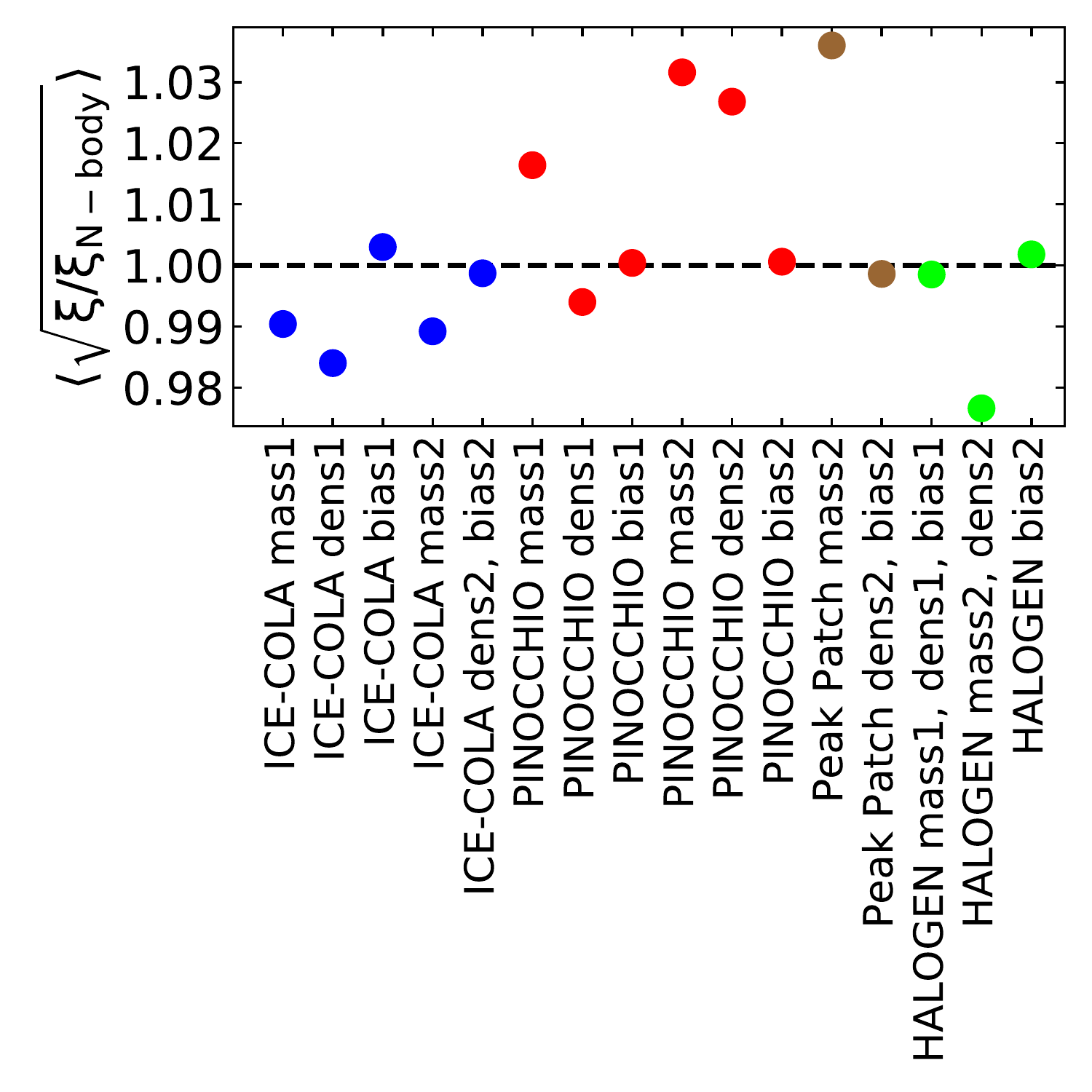}
 \end{minipage}
  \caption{Ratios of the total halo number (left panel) and the clustering amplitude 
  (right panel) of samples drawn from the approximate methods to 
  the corresponding quantity in the Minerva parent samples.
  By definition, for the dens samples the halo number is matched to the corresponding N-body samples and therefore the corresponding ratio is close to one in the left panel, while the ratios from the bias samples are meant to be close to one in the right panel. For some cases two or three samples are represented with the same symbol, e.g. \cola dens2, bias2 which means that the \cola dens2 sample is the same as the \cola bias2 sample.}
  \label{fig:nhalos_bias_ratios}
\end{figure*} 

We define two parent halo samples from the Minerva simulations by selecting halos with 
masses $M\ge 1.12\times 10^{13} h^{-1}M_{\sun}$ and 
$M\ge 2.667\times 10^{13} h^{-1}$M$_{\sun}$,
corresponding to 42 and 100 dark matter particles, respectively. 
We apply the same mass cuts to the catalogues produced by the approximated 
methods included in our comparison. We refer to the resulting samples 
as ``mass1'' and ``mass2''.

Note that the \patchy and log-normal catalogues do not have mass information for 
individual objects and match the number density and bias of the 
parent samples from Minerva by construction. 
The Gaussian model predictions are also computed for the specific clustering amplitude
and number density as the mass1 and mass2 samples. 
For the other approximate methods, the samples obtained by applying these mass 
thresholds do not reproduce the clustering and the shot noise of the corresponding 
samples from Minerva. 
These differences are in part caused by the different applied methods for identifying or assigning halos, e.g. \peak uses spherical overdensities in Lagrangian space to define halo masses while most other methods are closer to FoF masses, as described in Section \ref{sec:methods}.
Therefore, for the \cola, \halogen, \peak and \pin
catalogues we also define samples by matching number density and clustering amplitude of 
the halo samples from Minerva.
For the number-density-matched samples, we find the mass cuts where the total number of 
halos in the samples drawn from each approximate method
best matches that of the two parent Minerva samples. We refer to
these samples as ``dens1'' and ``dens2''. 
Analogously, we define bias-matched samples by identifying the mass thresholds for 
which the clustering amplitude in the catalogues drawn from the approximate 
methods best agrees with that of the mass1 and mass2 samples from Minerva.
More concretely, we define the clustering-amplitude-matched samples 
by selecting the mass thresholds that minimize the difference
between the mean correlation function measurements from the catalogues 
drawn from the approximate methods and the Minerva parent samples on 
scales $40\,h^{-1}\,{\rm Mpc}<s<80\,h^{-1}\,{\rm Mpc}$. 
We refer to these samples as ``bias1'' and ``bias2''.

 The mass thresholds defining the different samples, the number of particles
corresponding to these limits, their halo number densities, and bias ratios with 
 respect to the Minerva parent samples are listed in Table~\ref{tab:numden_limits}.
Note that, as the halo masses of the \pin and \peak catalogues 
are made continuous for this analysis, the mass cuts defining the density- and bias-matched samples 
do not correspond to an integer number of particles. 
Also note for the calibrated methods that the \halogen catalogue was calibrated using the input HMF from the mean of the 
300 Minerva simulations in logarithmic mass bins for this analysis, whereas the \patchy mass samples were calibrated for each mass cut individually.  For the case of the \halogen catalogue, the selected high mass threshold lies nearly half way (in logarithmic scale) between two of the mass thresholds of the logarithmic input HMF. This explains why whereas for the first mass cut, bias and number density are matched by construction, that is not the case for the second mass cut. This has the effect that the bias2 sample of the \halogen catalogue has 15\% fewer halos than the corresponding Minerva sample.
Comparisons of the ratios of the number densities and bias of the different samples 
drawn from the approximate methods to the corresponding ones from Minerva are shown 
in Fig.~\ref{fig:nhalos_bias_ratios}.
Since the catalogues drawn from log-normal and \patchy match the number density 
and bias of the Minerva parent samples by construction, they are not included in the 
Table and figures.

In the following we refer to all samples corresponding to the first mass limit, mass1, dens1 and bias1, as ``sample1'', and the samples corresponding to the second mass limit, mass2, dens2 and bias2, as ``sample2''.

\begin{table*}
	\centering
	\caption{Overview of the different samples, including the mass limits, $M_{\rm lim}$, 
	expressed in units of $h^{-1}{\rm M}_{\sun}$, the corresponding number of 
	particles, $N_{\rm p}$, the mean number density, $\bar{n}$, and the bias ratio to 
	the corresponding Minerva parent sample, 
	$\langle\left(\xi _{\rm app}/\xi _{Min}\right)^{1/2}\rangle$.
	The sample names ``mass'', ``dens'', and ``bias'', indicate if the samples were 
	constructed by matching the mass threshold, number density, or clustering amplitude 
	of the parent halo samples from Minerva.}
	\label{tab:numden_limits}
	\begin{tabular}{llllll}
		\hline
		code & sample name & $M_{\rm lim}/\left(h^{-1}{\rm M}_{\sun}\right)$ & $N_p$ & $\bar{n}/\left(h^3{\rm Mpc}^{-3}\right)$ & bias ratio   \\		
		\hline
Minerva & mass1 & $1.12\times 10^{13}$ & 42 & $2.12\times 10^{-4}$ & 1.00 \\
Minerva & mass2 & $2.67\times 10^{13}$ & 100 & $5.42\times 10^{-5}$  & 1.00 \\
\cola & mass1 & $1.12\times 10^{13}$ & 42 & $2.06\times 10^{-4}$  & 0.99 \\               
\cola & dens1 & $1.09\times 10^{13}$ & 41 & $2.12\times 10^{-4}$ & 0.98\\
\cola & bias1 & $1.17\times 10^{13}$ & 44 & $1.93\times 10^{-4}$ & 1.00 \\
\cola & mass2 & $2.67\times 10^{13}$ & 100 & $5.81\times 10^{-5}$ & 0.99\\  
\cola & dens2, bias2 & $2.77\times 10^{13}$ & 104 & $5.45\times 10^{-5}$ & 1.00 \\
\halogen & mass1, dens1, bias1 & $1.12\times 10^{13}$ & 42 & $2.14\times 10^{-4}$ & 1.00\\
\halogen & mass2, dens2 & $2.67\times 10^{13}$ & 100 & $5.40\times 10^{-5}$ & 0.98 \\
\halogen & bias2 & $2.91\times 10^{13}$ & 109 & $4.61\times 10^{-5}$ & 1.00 \\
  \peak \footnotemark & mass2 & $2.67\times 10^{13}$ & 100 & $4.45\times 10^{-5}$  & 1.04\\                
\peak & dens2, bias2 & $2.35\times 10^{13}$ & 88.3 & $5.44\times 10^{-5}$ & 1.00\\
\pin & mass1 & $1.12\times 10^{13}$ & 42 & $1.95\times 10^{-4}$ & 1.02 \\
\pin & dens1 & $1.04\times 10^{13}$ & 39.1 & $2.15\times 10^{-4}$ & 1.00\\
\pin & bias1 & $1.06\times 10^{13}$ & 39.9 &  $2.09\times 10^{-4}$ & 1.00 \\
\pin & mass2 & $2.67\times 10^{13}$ & 100 & $5.35\times 10^{-5}$ & 1.03\\
\pin & dens2 & $2.63\times 10^{13}$ & 98.6 &  $5.48\times 10^{-5}$ & 1.03\\
\pin & bias2 & $2.42\times 10^{13}$ & 90.7 & $6.27\times 10^{-5}$	 & 1.00\\
	\end{tabular}
\end{table*}
\footnotetext{As the halo masses corresponding to our low-mass threshold are not correctly 
resolved in the \peak catalogues, only the high-mass threshold (mass2) is considered in this case.}

\subsection{Clustering measurements in configuration space}
\label{sec:2pcf}

Most cosmological analyses of galaxy redshift surveys are based on 
two-point clustering statistics. In this paper we focus on 
configuration-space analyses and study the estimation of the 
covariance matrix of correlation function measurements. 
The information of the full two-dimensional correlation function, $\xi(s,\mu)$, where 
$\mu$ is the cosine of the angle between the separation vector ${\bf s}$ and the line
of sight, can be compressed into a small number of functions such as 
the Legendre multipoles, $\xi_\ell(s)$, given by 
\begin{equation}
 \xi_\ell(s) = \frac{2\ell+1}{2} \int _{-1} ^1 L_\ell(\mu)\xi(\mu,s) \dint \mu,
 \label{eq:xi_multi}
\end{equation}
where $L_\ell(\mu)$ denotes the Legendre polynomial of order $\ell$. Typically, only
multipoles with $\ell \leq 4$ are considered.
An alternative tool is the clustering wedges statistic \citep{Kazin2012}, which 
corresponds to the average of the full two-dimensional correlation function over 
wide bins in $\mu$, that is
\begin{equation}
 \xi _{\mathrm{w},i} (s) = \frac{1}{\Delta \mu} \int _{(i-1)/n} ^{i/n} \xi(\mu,s) \dint \mu,
 \label{eq:xi_wedges}
\end{equation}
where $\xi _{\mathrm{w},i} $ denotes each individual clustering wedge, and $n$ 
represents the total number of wedges. We follow the recent analysis of 
\citet{Sanchez2017} and divide the $\mu$ range from 0 to 1 into three equal-width 
intervals, $i=1,2,3$. 

We compute the Legendre multipoles and clustering wedges of the halo samples defined
in Section~\ref{sec:samples}.
As these measurements are based on simulation boxes with periodic boundary conditions, 
the full $\xi(s,\mu)$ can be computed using the natural estimator, namely
\begin{equation}
\xi(s,\mu)=\frac{DD(s,\mu)}{RR(s,\mu)}-1,
\end{equation}
where $DD(s,\mu)$ are the normalized data pair counts and $RR(s,\mu)$ the normalized 
random pair counts, which can be computed as the ratio of the volume of a shell $dV$ and 
the total box volume $V_{\rm s}$, $RR = dV/V_{\rm s}$.
The obtained $\xi(s,\mu)$ can be used to estimate Legendre multipoles and clustering wedges 
using equations (\ref{eq:xi_multi}) and (\ref{eq:xi_wedges}), respectively. 
We consider scales in the range 
$20\,\,h^{-1}{\rm Mpc} \leq s \leq 160\,\,h^{-1}{\rm Mpc}$  for all our measurements and 
implement a binning scheme with $ds=10\,h^{-1}{\rm Mpc}$ for the following analysis. 
For illustration purposes we also use a binning of $ds=5\,h^{-1}{\rm Mpc}$ for the figures showing correlation function measurements.
Considering Legendre multipoles with $\ell \leq 4$ and three $\mu$ wedges, the dimension of the 
total data vector, $\bm{\xi}$, containing all the measured statistics is the same in 
both cases ($N_{\rm b} = 42$ and $N_{\rm b} = 84$ for the cases of 
$ds=10\,h^{-1}{\rm Mpc}$ and $ds=5\,h^{-1}{\rm Mpc}$, respectively).

\begin{figure*}

 \includegraphics[width=\textwidth]{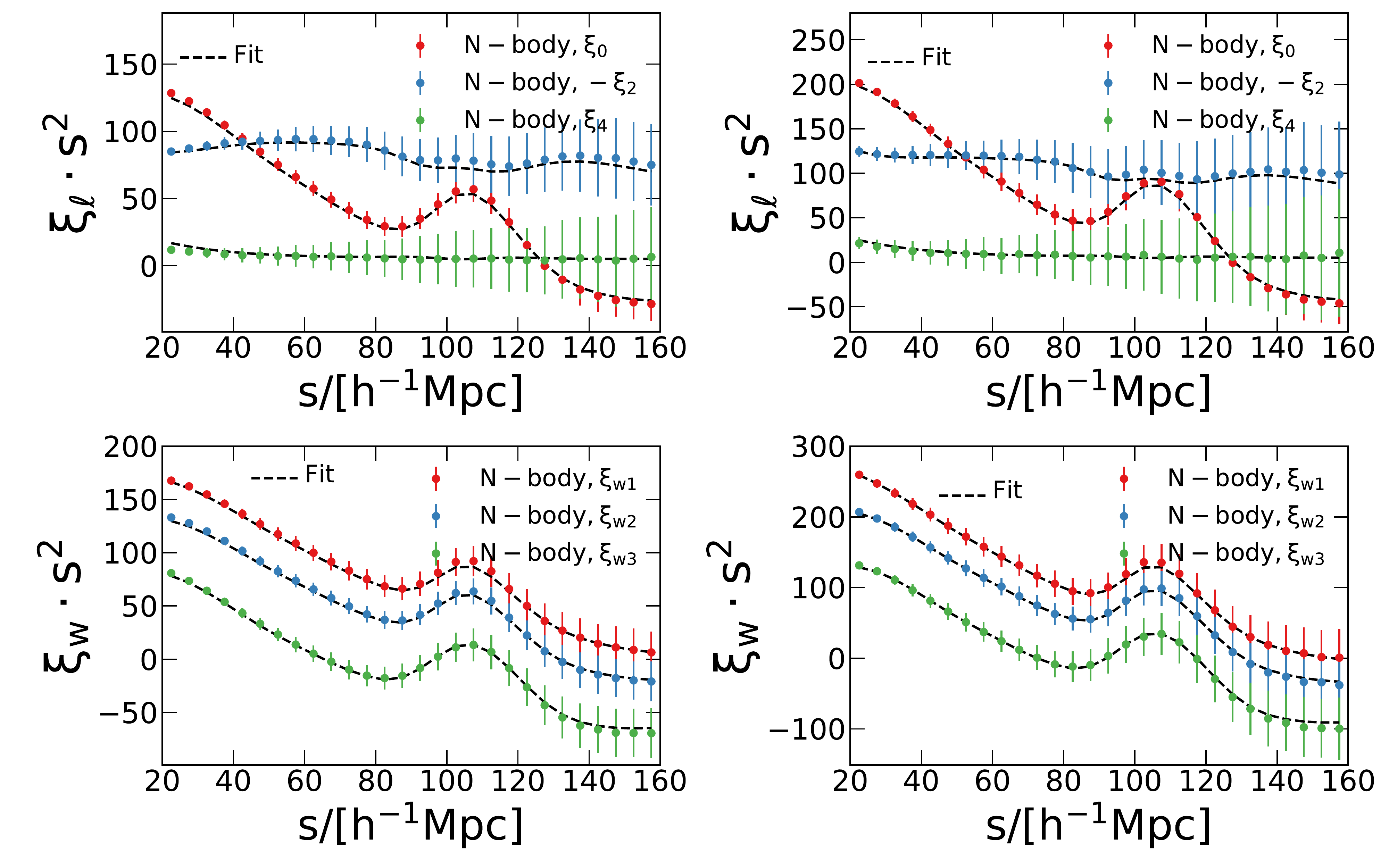}
  \caption{Comparison of the mean correlation function multipoles (upper panels) and 
  clustering wedges (lower panels) of the mass1 and mass2  samples (left and right panels, 
  respectively) drawn from our Minerva N-body simulations, and the model described in Section~\ref{sec:xi_model}.
  The points with error bars show to the simulation results and the dashed lines 
  correspond to the fit to these measurements. The error bars on the measurements 
  correspond to the dispersion inferred from the 300 Minerva realizations. In all 
  cases, the model predictions show good agreement with the N-body measurements.}
  \label{fig:minerva_fit}
\end{figure*}

\subsection{Covariance matrix estimation}
\label{sec:cov-mat-est}

It is commonly assumed that the likelihood function of the measured two-point 
correlation function is Gaussian in form, 
\begin{equation}
 -2\ln \mathcal{L} (\bm{\xi}|\bm{\theta}) = (\bm{\xi}-\bm{\xi_\mathrm{theo}}(\bm{\theta}))^t\bm{\Psi} (\bm{\xi}-\bm{\xi_\mathrm{theo}}(\bm{\theta}))
 \label{eq:likelihood}
\end{equation}
where $\bm{\xi_\mathrm{theo}}$ represents the theoretical model of the measured 
statistics, which here correspond to the Legendre multipoles or clustering wedges, 
for the parameters $\bm{\theta}$, and $\bm{\Psi}$ is the precision matrix,  
given by  the inverse of the covariance matrix, $\bm{\Psi} = \mathbfss{C}^{-1}$.

The covariance matrix, $\mathbfss{C}$ is usually estimated from a large set of $N_{\rm s}$ 
mock catalogues as
\begin{equation}
 C_{ij} = \frac{1}{N_{\rm s} -1} \sum _{k=1} ^{N_{\rm s}} (\xi_i ^k -\bar{\xi_i})(\xi_j ^k -\bar{\xi_j}),
 \label{eq:cov_est}
\end{equation}
where $\bar{\xi_i} = \frac{1}{N_{\rm s}}\sum _k \xi _i ^k$ is the mean value of 
the measurements at the $i$-th bin and $\xi_i ^k$ is the corresponding measurement from the $k$-th mock. 
This estimator has the advantage 
over other techniques such as Jackknife estimates from the data or theoretical 
modelling, that it tends to be less affected by biases than estimates from the data and does not require any assumptions regarding the 
properties of the true covariance matrix.
However, the noise in $\mathbfss{C}$ due to the finite number of realizations 
leads to an additional uncertainty, which must be propagated into the final parameter 
constraints \citep{Taylor2013, Dodelson2013,Percival2014, Sellentin2016}.
Depending on the analysis configuration, the control of this additional error might require 
a large number of realizations, with $N_{\rm s}$ in the range of a few thousands.
For the new generation of large-volume surveys such as {\it Euclid}, the construction of
a large number of mock catalogues might be extremely demanding and will need to rely, 
at least partially, on approximate N-body methods. The goal of our analysis is to test
the impact on the obtained parameter constraints of using estimates of $\mathbfss{C}$
based on different approximate methods. 

We use equation~(\ref{eq:cov_est}) to compute the covariance matrices associated with 
the measurements of the multipoles and clustering wedges of the halo samples defined in 
Section~\ref{sec:samples}.
In order to reduce the noise in these measurements due to the limited number of 
realizations, we obtain three separate estimates of $\mathbfss{C}$ from each 
sample by treating each axis of the simulation boxes as the line-of-sight direction
when computing $\xi(s,\mu)$. 
Our final estimates correspond to the average of 
the covariance matrices measured on the different lines of sight.
The Gaussian theoretical covariance matrices were computed for the specific number 
density and clustering of the halo samples from Minerva. We used as input the model of the 
two-dimensional power spectrum described in Section~\ref{sec:xi_model}, whose 
parameters were fitted to reproduce the clustering of parent halo samples.  

\subsection{Testing the impact of approximate methods for covariance matrix estimates}
\label{sec:xi_model}

The cosmological information recovered from full-shape fits to anisotropic 
clustering measurements is often expressed in terms of the BAO shift parameters 
\begin{align}
\alpha_\perp  &= \frac{D_{\rm A}(z)\,r_{\rm d}'}{D_{\rm A}'(z)\,r_{\rm d}}, \\
\alpha_{\parallel} &= \frac{H'(z)\,r_{\rm d}'}{H(z)\,r_{\rm d}},
\end{align}
where $H(z)$ is the Hubble parameter at redshift $z$, $D_{\rm A}(z)$ is the corresponding 
angular diameter distance, $r_{\rm d}$ is the sound horizon at the drag redshift,
and the primes denote quantities in the fiducial cosmology; 
and the RSD parameter combination $f\sigma_8(z)$, where $f(z)$ represents the 
logarithmic growth rate of density fluctuations and $\sigma_8(z)$ is the linear rms mass fluctuation in spheres of 
radius $8\,h^{-1}{\rm Mpc}$. 

The constraints on these parameters are sensitive to details in the definition 
of the likelihood function, such as the way in which the covariance matrix of
the measurements is estimated. In order to asses the impact of using approximate 
methods to estimate $\mathbfss{C}$, we perform full-shape fits of anisotropic 
clustering measurements in configuration space to obtain constraints on 
$\alpha_\perp$, $\alpha_{\parallel}$, and $f\sigma_8(z)$ assuming the 
Gaussian likelihood function of equation~(\ref{eq:likelihood}). 
We compare the constraints obtained when $\mathbfss{C}$ is estimated from a set 
of full N-body simulations with the results inferred from the same set of 
measurements when the covariance matrix is computed using the approximate 
methods described in Sec.~\ref{sec:methods}.

\begin{figure*}
\begin{minipage}[c]{\textwidth}
\centering
 \includegraphics[width=0.64\textwidth]{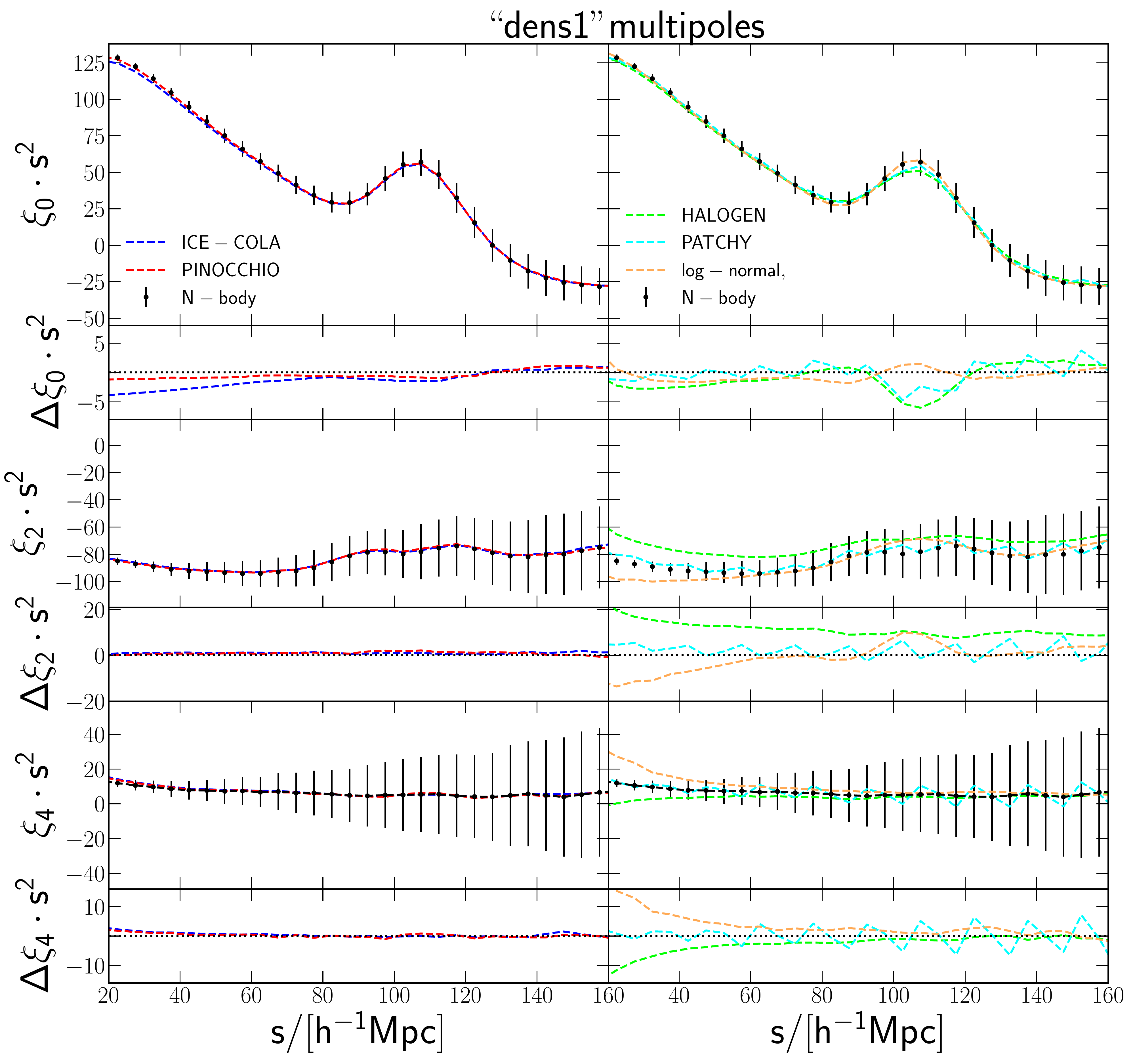}
\end{minipage}
\begin{minipage}[c]{\textwidth}
\centering
 \includegraphics[width=0.64\textwidth]{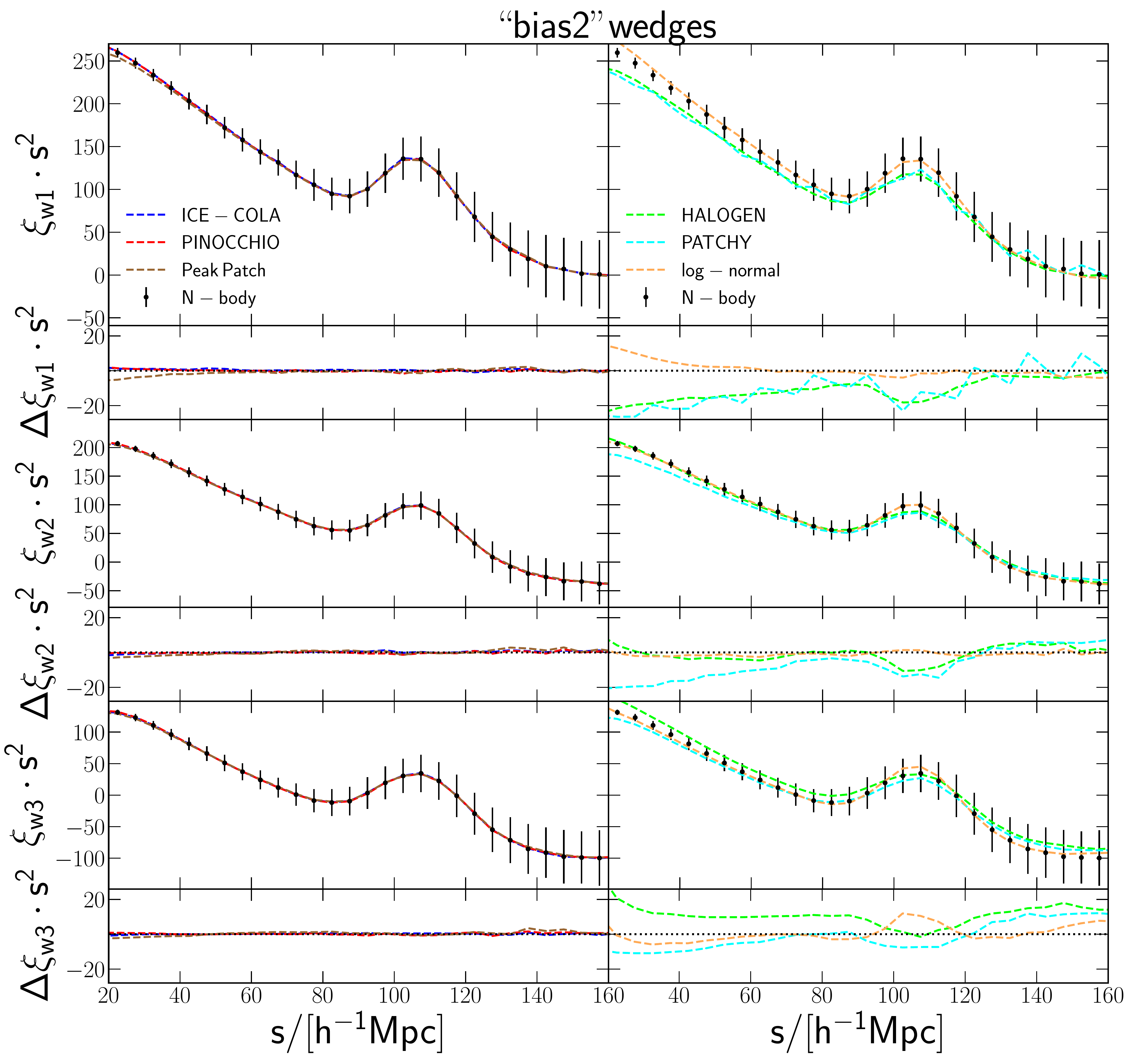}
 \end{minipage}
 \caption{\textit{Upper panel}: measurements of the mean multipoles for the density 
matched samples for the first mass cut (dens1 samples). The first, third and fifth 
row show the monople, quadrupole and hexadecapole, respectively. \textit{Lower panel}: 
measurements  of the mean clustering wedges for the bias matched samples for the second 
mass cut (bias2 samples).  The first, third and fifth row show the transverse, 
intermediate and parallel wedge, respectively. Comparison of the measurements drawn from 
the results of the predictive methods \cola and \pin (\textit{left panels}) and the 
calibrated methods \halogen and \patchy and the log-normal model (\textit{right panels})
to the corresponding N-body parent sample. The error bars correspond to the dispersion 
of the results inferred from the 300 N-body catalogues. The remaining rows show the 
difference of the mean measurements drawn from the results of the approximate methods 
to the corresponding N-body measurement.}
 \label{fig:xi_comp}
\end{figure*}

Our fits are based on the same model of the full two-dimensional correlation 
function $\xi(\mu,s)$ as in the analyses of the final 
BOSS galaxy samples \citep{Sanchez2017, Grieb2017,Salazar-Albornoz2017}, 
and the eBOSS DR12 catalogue \citep{Hou2018}.
This model includes the effects of the non-linear evolution of density 
fluctuations based on gRPT (Crocce et al . in prep.) 
bias \citep{Chan2012}, and 
redshift-space distortions (Scoccimarro et al. in prep.).
The only difference between the model implemented in these studies 
and the one used here is that, 
since we analyse halo samples instead of galaxies, we do not include the
so-called fingers-of-God factor, $W_\infty(k,\mu)$ 
\citep[see equation 18 in][]{Sanchez2017}. 
In total, our parameter space contains six free parameters, the BAO and RSD 
parameters $\alpha_{\parallel}$, $\alpha_{\perp}$, and  $f\sigma_8$, 
and the nuisance parameters associated with the linear and quadratic local bias, 
$b_1$ and $b_2$, and the non-local bias $\gamma_3^-$.
We explore this parameter space by means of the Monte Carlo Markov Chain (MCMC)
technique.
This analysis set-up matches that of the covariance matrix comparison in Fourier 
space of our companion paper \citet{Blot18}.

In order to ensure that the model used for the fits has no impact on the 
covariance matrix comparison, we do not fit the measurements of the Legendre 
multipoles and wedges obtained from the N-body simulations. Instead, we 
use our baseline model to construct synthetic clustering measurements, 
which we then use for our fits.
For this, we first fit the mean Legendre multipoles measured from the parent 
Mineva halo samples using our model and the N-body covariance matrices.
We fix all cosmological parameters to their true values and only vary the bias 
parameters $b_1$, $b_2$, and $\gamma_3^-$.
We then use the mean values of the parameters inferred from the fits, together 
with the true values of the cosmological parameters, to generate multipoles and 
clustering wedges of the correlation function using our baseline model.
Fig.~\ref{fig:minerva_fit} shows the mean multipoles and clustering wedges measured 
from the Minerva halo sample for both mass cuts and the resulting fits. 
In all cases, our model gives a good description of the simulation results.
The parameter values recovered from these fits were also used to compute the 
input power spectra 
when computing the Gaussian predictions of $\mathbfss{C}$. 
As these synthetic data are perfectly described by our baseline model by construction,
their analysis should recover the true values of the BAO parameters 
$\alpha_{\parallel} = \alpha_{\perp} = 1.0$, and the 
growth-rate parameter $f\sigma_8=0.4402$.
The comparison of the parameter values and their uncertainties recovered 
using different covariance matrices allows us to test the ability of the 
approximate methods described in Section~\ref{sec:methods} to reproduce 
the results obtained when $\mathbfss{C}$ is inferred from full N-body simulations.

\section{Results}
\label{sec:results}

In this section, we present a detailed comparison of the covariance matrix measurements 
in configuration space obtained from the approximate methods described in 
Section~\ref{sec:methods} and their performance at recovering the correct parameter 
estimates.

\subsection{Two-point correlation function measurements}
\label{sec:cf-comp}

In order to estimate the covariance matrices from all the samples introduced in 
Section~\ref{sec:samples}, we first measure configuration-space Legendre multipoles 
and clustering wedges for each sample and in each realization as described in 
Section~\ref{sec:2pcf}. 

As an illustration of the agreement between the clustering measurements obtained 
from the approximate methods and the Minerva simulations we focus here on two
cases: i) the multipoles of the density-matched samples for the first mass cut 
(dens1 samples), and  ii) the clustering wedges of the bias-matched 
samples for the second mass cut (bias2 samples). 
As described in Section~\ref{sec:samples}, for \patchy and the log-normal 
realizations, the density- and bias-matched samples are identical to the 
mass-matched samples by construction.

The upper panel of Fig.~\ref{fig:xi_comp} shows the mean multipole measurements 
from all realizations for the dens1 samples obtained from the predictive methods \cola 
and \pin (left panels) and the calibrated methods \halogen, \patchy, and the 
log-normal recipe (right panels). 
The predictive methods are in excellent agreement with the measurements from the 
Minerva parent sample, showing only differences of less than 3\% for the \cola 
monopole measurements on scales $< 40\,h^{-1}$Mpc. The monopole measurements obtained 
from the calibrated methods and the log-normal model are also in good agreement with 
the results from Minerva. However, the quadrupole and hexadecapole measurements 
obtained from \halogen and the log-normal samples exhibit deviations of more than 
20\% on scales $< 60\,h^{-1}{\rm Mpc}$.

The lower panel of Fig.~\ref{fig:xi_comp} shows the mean wedges measurements 
from all realizations for the bias2 samples obtained from the predictive methods \cola, 
\pin and \peak (left panels),  and for the corresponding samples obtained 
from calibrated methods \halogen, \patchy, and the log-normal recipe (right panels). 
Here we find that the measurements obtained from the predictive methods and the log-normal 
model agree well within the error bars with the corresponding Minerva measurements. We 
notice that the strongest deviations are present in the measurements of the transverse 
and parallel wedge from the \halogen samples, of up to 6\% and 20\% respectively on scales 
$< 60\,h^{-1}$Mpc. The measurements recovered from \patchy show deviations ranging 
between 5\% to 10\% on small scales.

\begin{figure}
\begin{minipage}{\columnwidth}
 \includegraphics[width=\columnwidth]{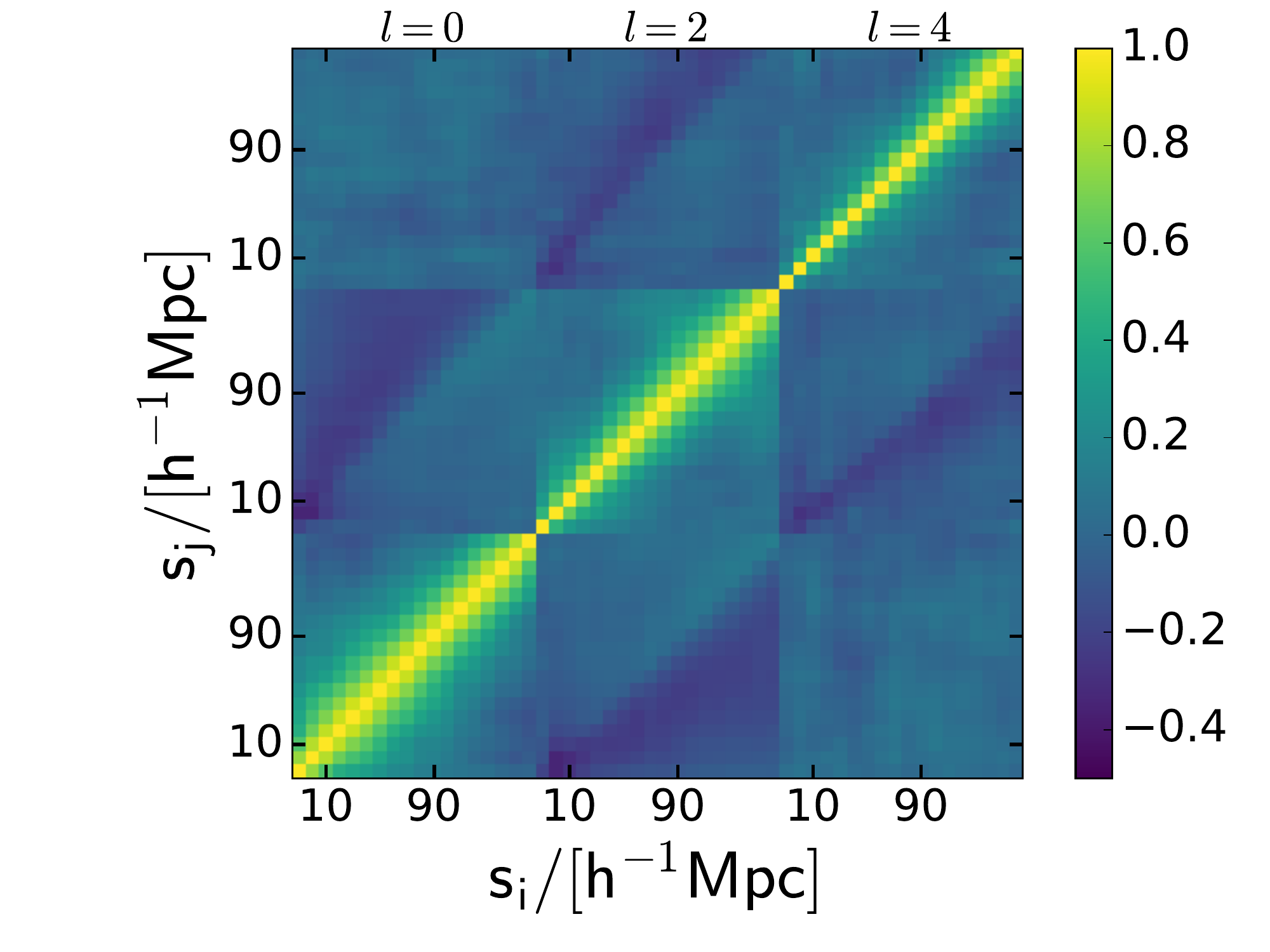}
  \includegraphics[width=\columnwidth]{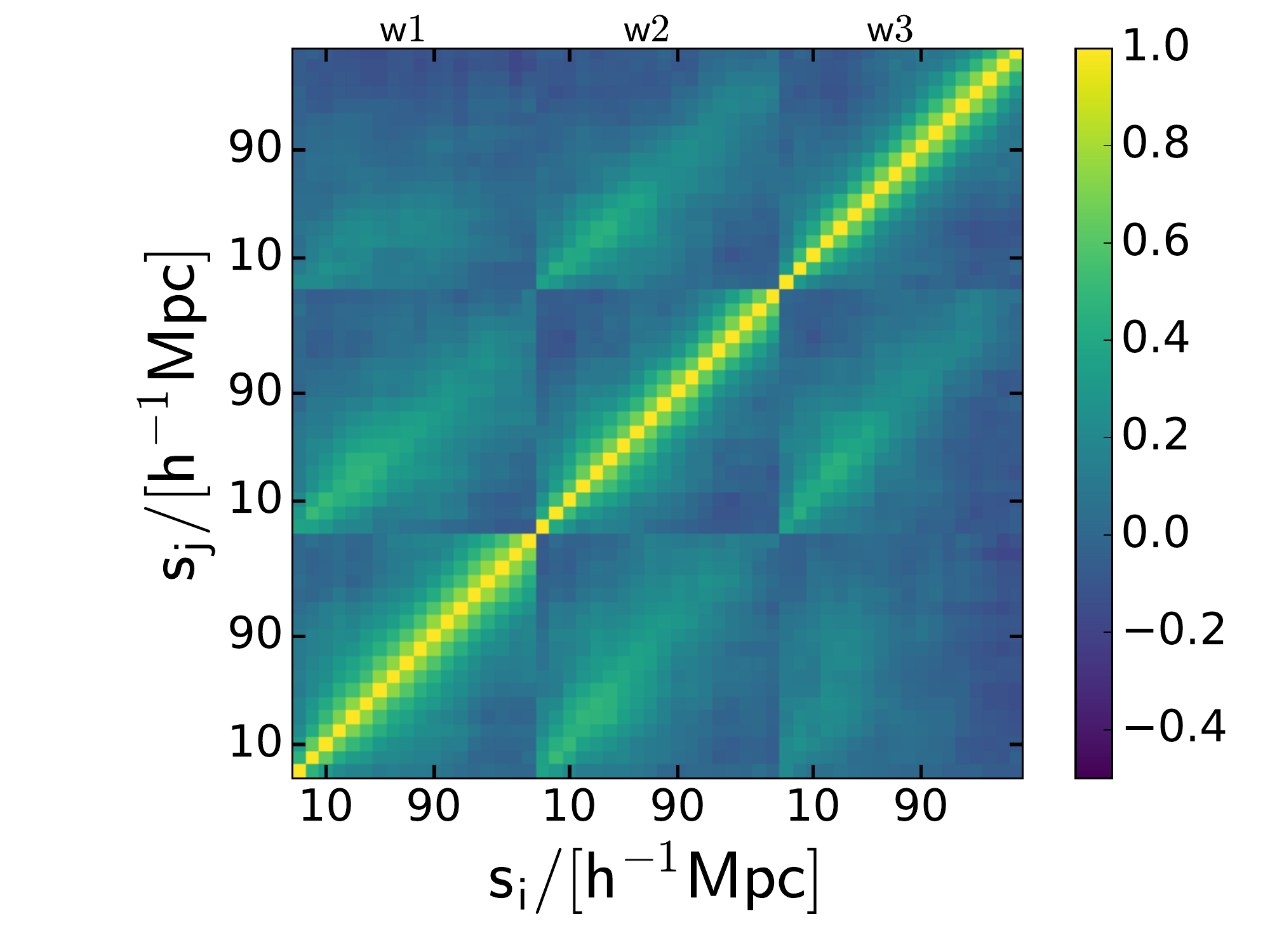}
 \end{minipage}
 \caption{The full correlation matrix inferred from the multipoles of the N-body parent sample 
 for the low-mass cut (mass1, upper panel) and from the clustering wedges of the mass2 N-body 
 parent sample (lower panel).}
 \label{fig:corr_minerva}
\end{figure}

\subsection{Covariance matrix measurements}
\label{sec:cov_mat_comp}

In this section we focus on the comparison of the covariance matrix estimates 
obtained from the different approximate methods, which we computed as described in 
Section~\ref{sec:cov-mat-est}.

The structure of the off-diagonal elements of $\mathbfss{C}$ 
of Legendre multipoles and clustering wedges measurements can be more clearly seen 
in the correlation matrix, defined as
\begin{equation}
R_{ij} = \frac{C_{ij}}{\sqrt{C_{ii}C_{jj}}}.
\label{eq:corr_mat}
\end{equation}
Fig.~\ref{fig:corr_minerva} shows the correlation matrices of the multipoles inferred 
from the mass1 halo samples from Minerva (upper panel) and the wedges of the mass2 
samples (lower panel). 

\begin{figure*}
\begin{minipage}[c]{\textwidth}
\centering
 \includegraphics[width=0.6\textwidth]{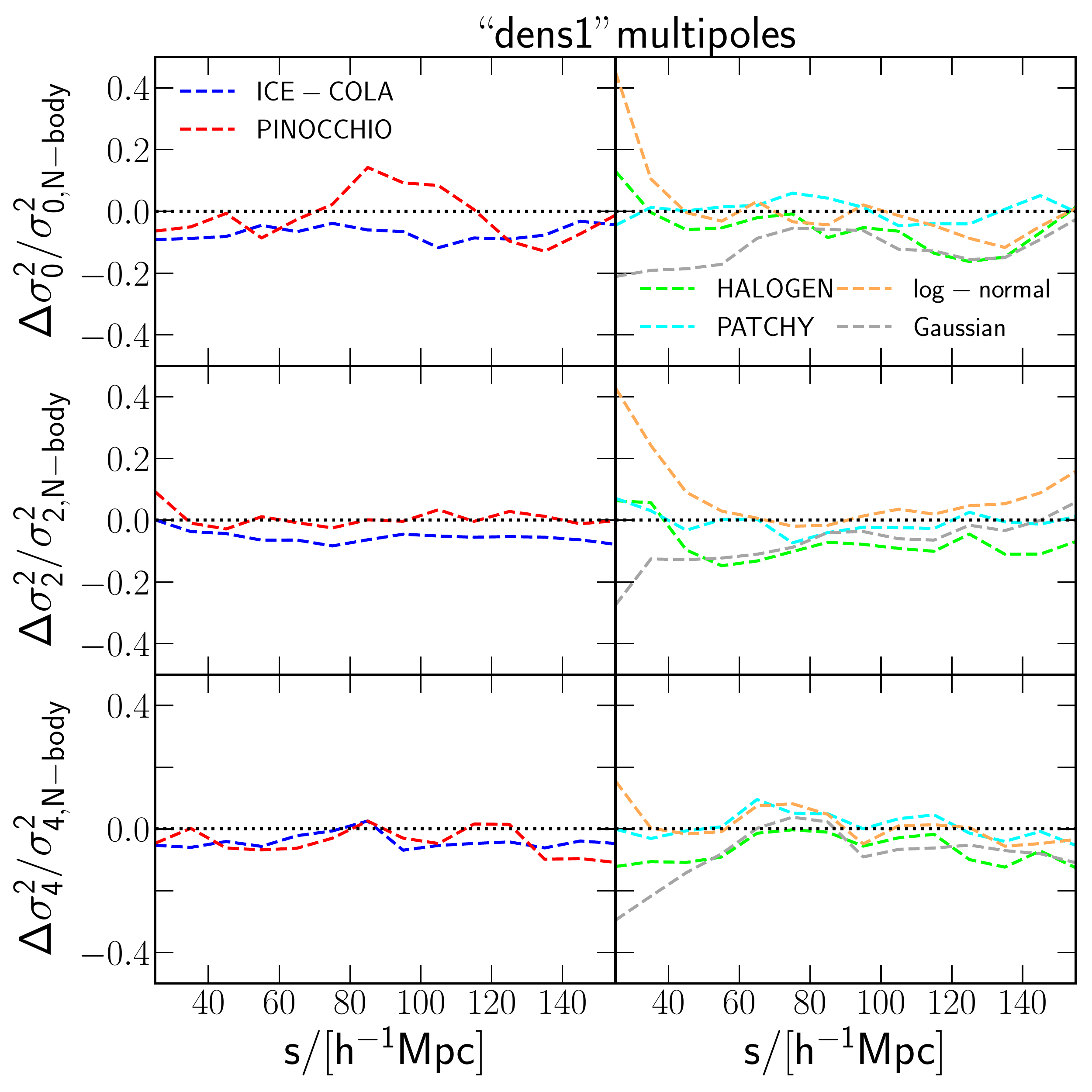}
\end{minipage}
\begin{minipage}[c]{\textwidth}
\centering
 \includegraphics[width=0.6\textwidth]{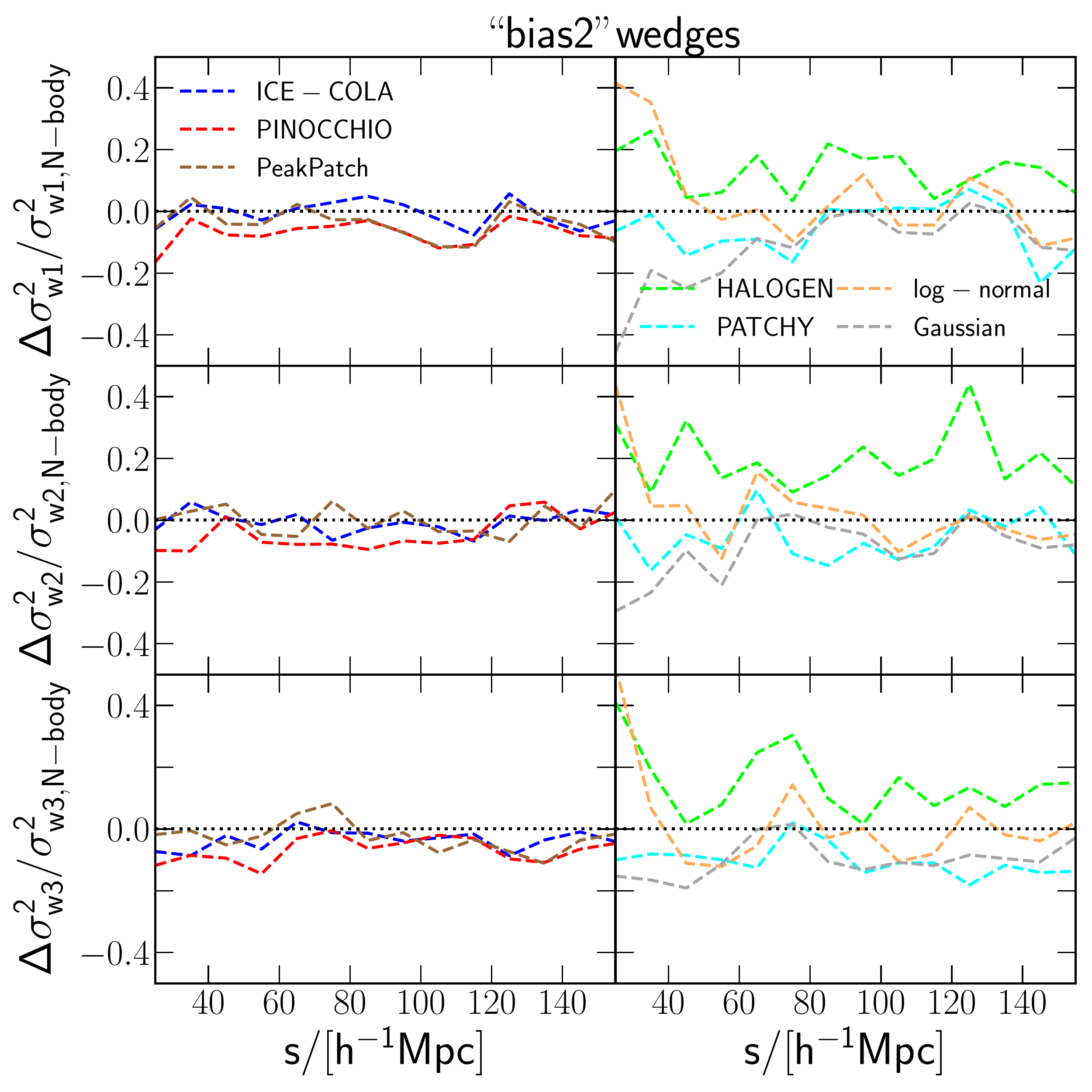}
 \end{minipage}
 \caption{ \textit{Upper panel}: Relative variance of the multipoles of the correlation 
function measurements from the density matched samples for the first mass cut (dens1 
samples).The first, third and fifth row show the measurements for monople, quadrupole 
and hexadecapole, respectively. \textit{Lower panel}: Relative variance of the clustering 
wedges of the two-point correlation function for the bias matched samples for the second 
mass cut (bias2 samples). The first, third and fifth row show the measurements 
for transverse, intermediate and parallel wedge, respectively.  Comparison of the 
relative variance drawn from the results of the predictive methods \cola, \pin, \peak (\textit{left panel}) and 
\halogen, \patchy and the 
log-normal model (\textit{right panel})  to the corresponding N-body parent sample.}
 \label{fig:var_comp}
\end{figure*}

The estimates of $\mathbfss{R}$ obtained from the approximate methods 
are indistinguishable by eye from the ones inferred from the Minerva 
parent samples and therefore not shown here.
Instead, we compare the variances and cuts through the correlation function matrices 
derived from the different samples.  
Fig.~\ref{fig:var_comp} shows the ratios of the variances  
drawn from the approximate methods with respect to those of the  
corresponding Minerva parent catalogues. We focus here on the same example cases 
as in Section~\ref{sec:cf-comp}: the multipoles measured from the dens1 samples, 
and the clustering wedges measured from the bias2 samples. 
We notice that in both cases the predictive methods 
perform better than the calibrated schemes and the PDF-based recipes. 
On average, the variance from Minerva is recovered within 10\%, with a maximum 
difference of 20\% for the variance of the monopole inferred from the 
\pin  dens1 sample at scales around $80\,h^{-1}{\rm Mpc}$. 
The variances recovered from the other methods show larger 
deviations, in some cases up to 40\%.

\begin{figure*}
\begin{minipage}[c]{\columnwidth}
\centering
 \includegraphics[width=\columnwidth]{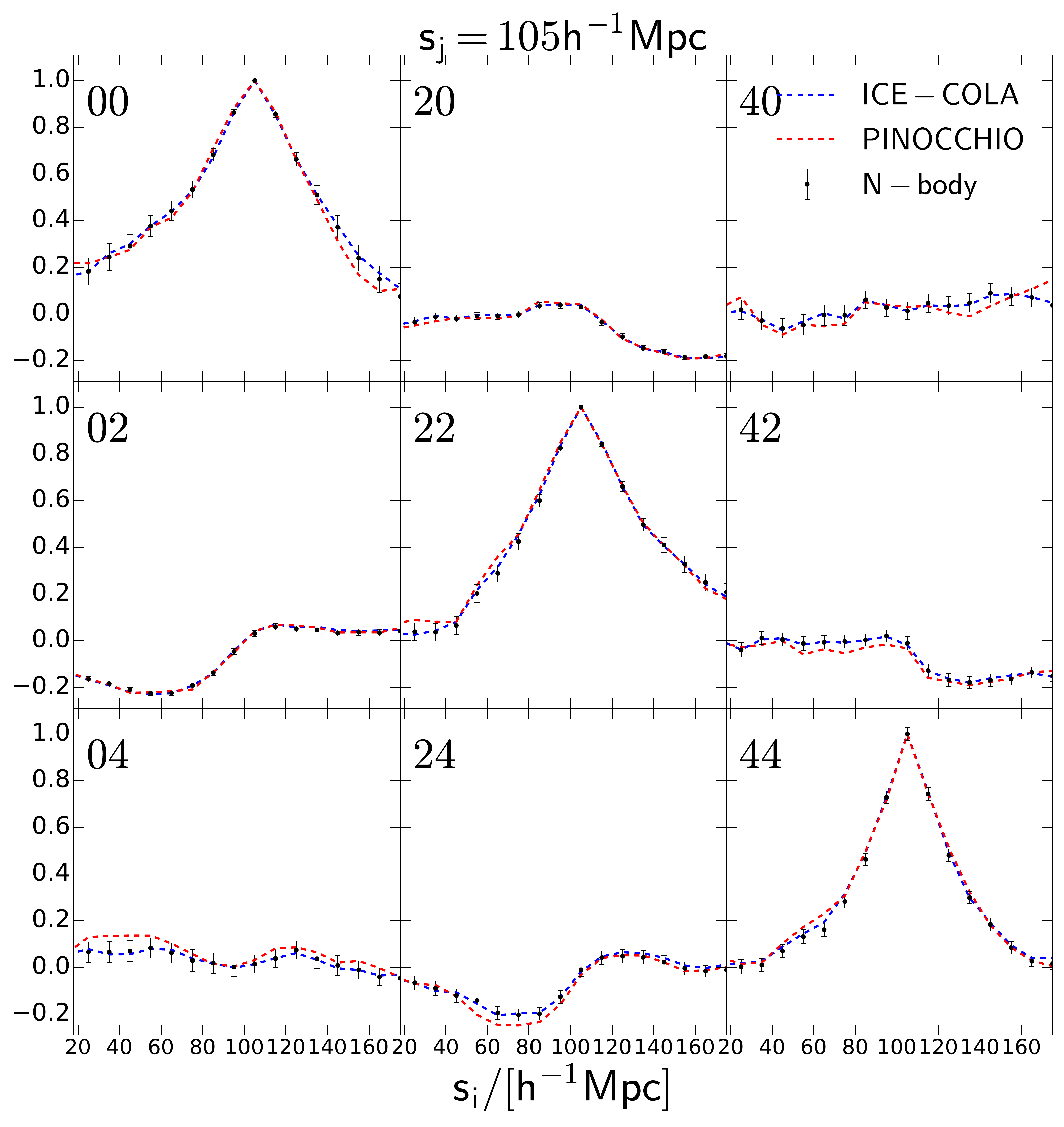}
 \includegraphics[width=\columnwidth]{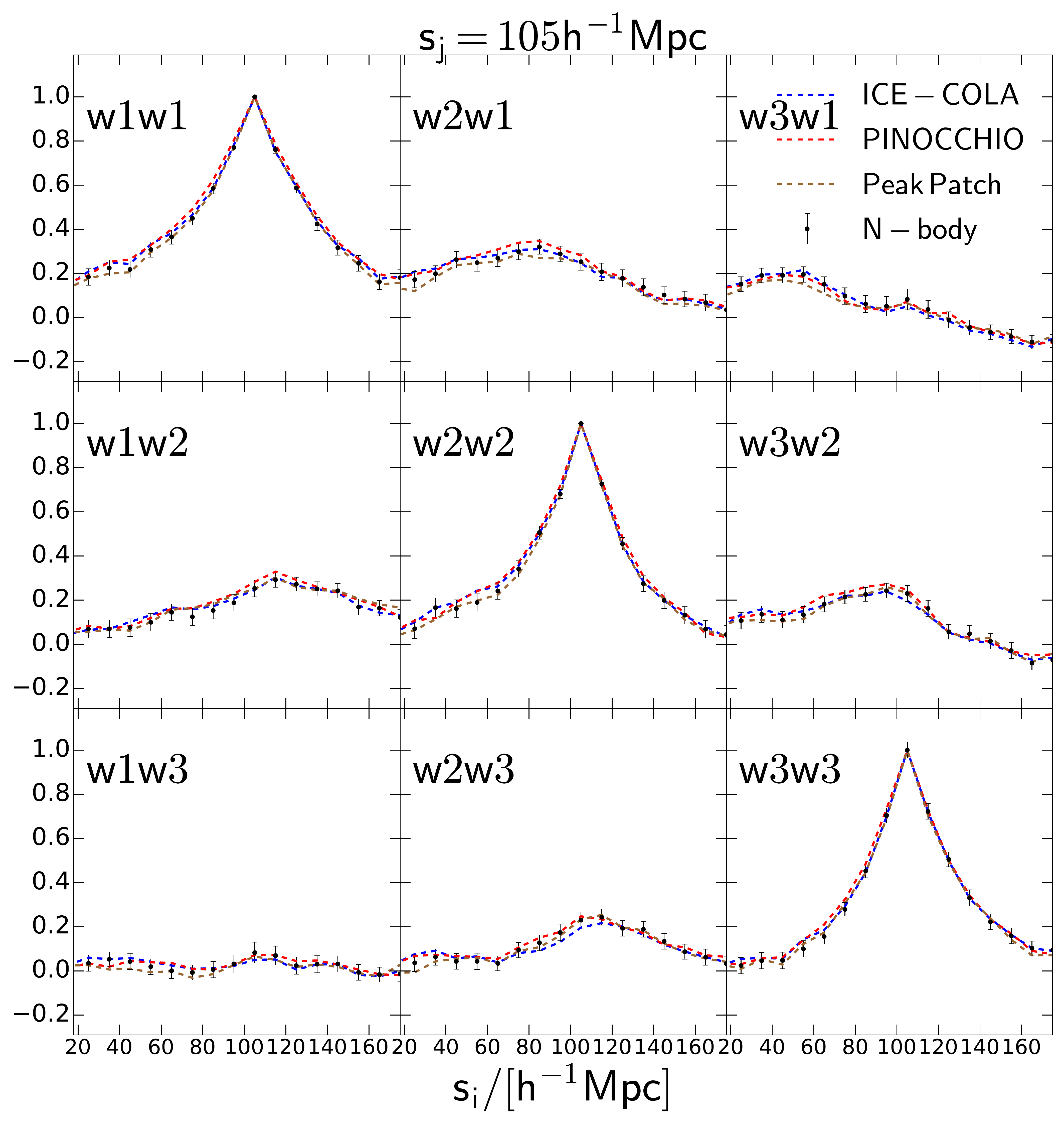}
\end{minipage}
\begin{minipage}[c]{\columnwidth}
\centering
 \includegraphics[width=\columnwidth]{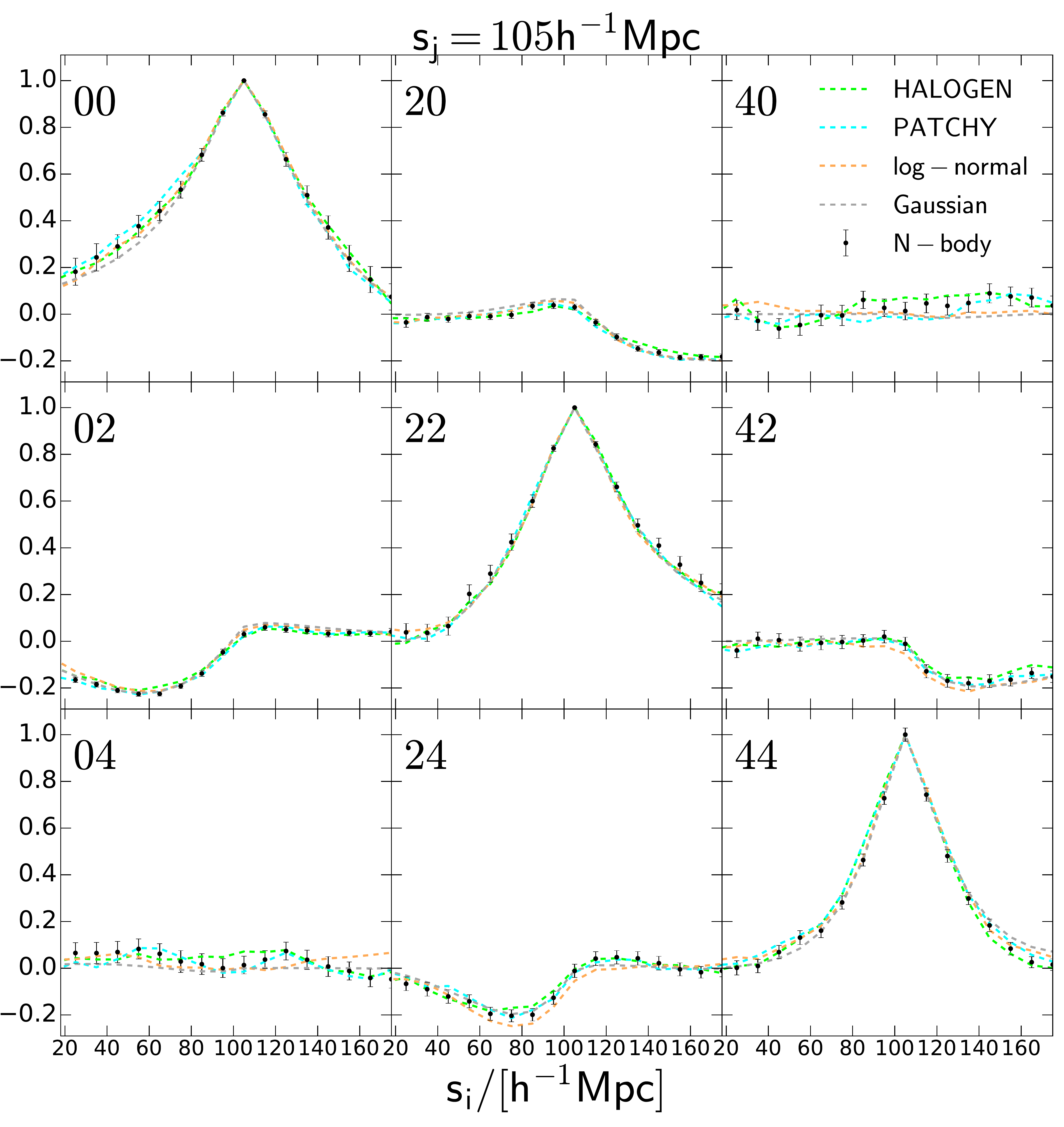}
 \includegraphics[width=\columnwidth]{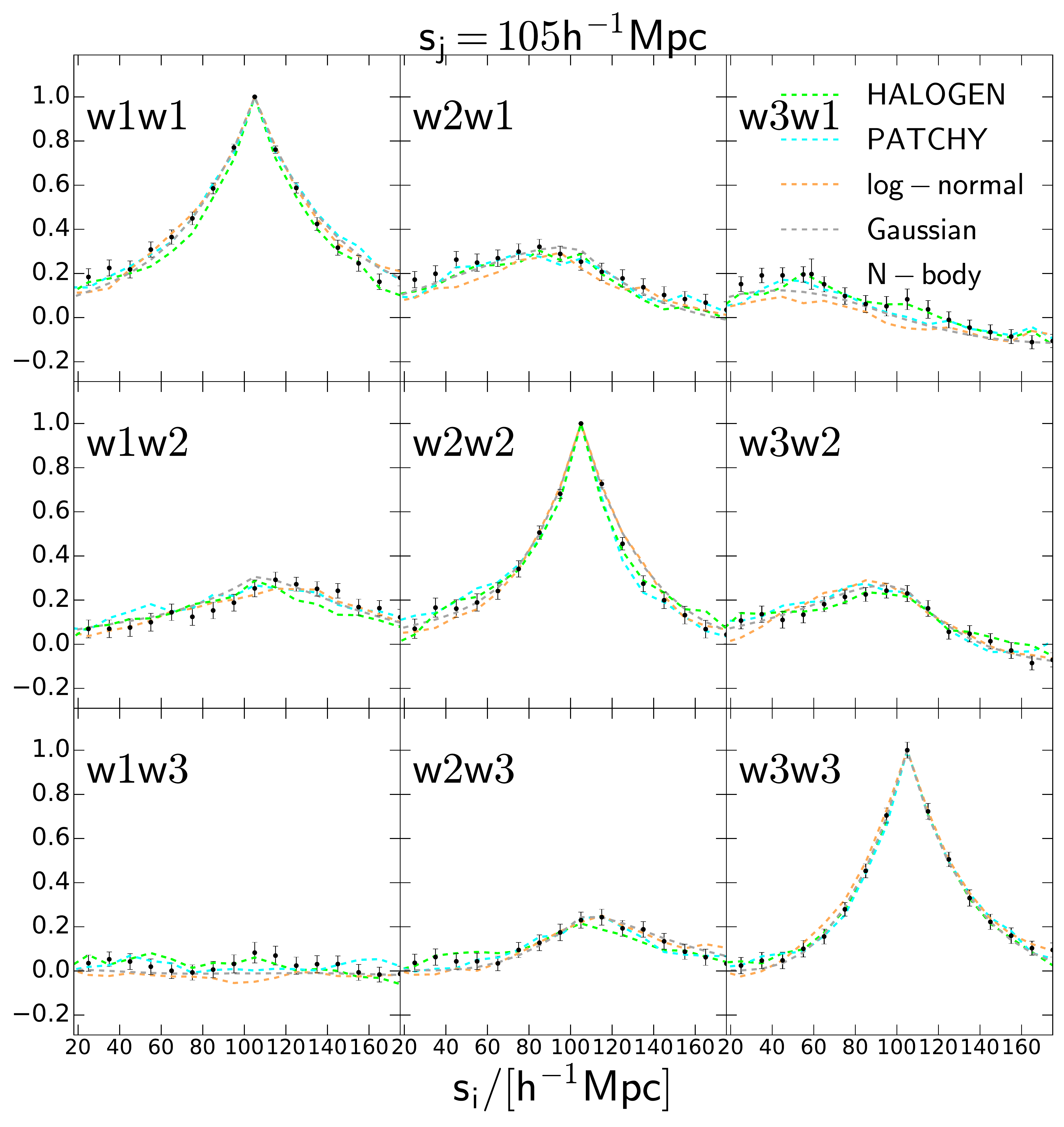}
 \end{minipage}
 \caption{Cuts at $s_j = 105\,h^{-1}$Mpc through the correlation matrices for the two 
 example cases drawn from the results of the approximate methods and the corresponding 
 N-body parent sample. 
 \textit{Upper, left panel}: Correlation matrices measured from the multipoles of the correlation function drawn from dens1 samples from the predictive methods \cola and \pin.
  \textit{Upper, right panel}: Correlation matrices measured from the multipoles of the correlation function drawn from dens1 samples from the calibrated methods \halogen and \patchy and the Gaussian and log-normal recipes.
   \textit{Lower, left panel}: Correlation matrices measured from the clustering wedges of the correlation function drawn from the bias2 samples from the predictive methods \cola, \pin and \peak.
   \textit{Lower, right panel}: Correlation matrices measured from the clustering wedges of the correlation function drawn from bias2 samples from the calibrated methods \halogen and \patchy and the Gaussian and log-normal recipes.
  The error bars are obtained from a jackknife estimate using the 300 Minerva realizations.}
 \label{fig:cut_corr}
\end{figure*}

Fig.~\ref{fig:cut_corr} shows cuts through the correlation 
matrix at $s_j = 105\,h^{-1}{\rm Mpc}$ for the same two example cases.
The error bars  for the measurements of the corresponding Minerva parent samples are obtained from a jackknife estimate using the 300 Minerva mocks,

\begin{equation}
(\Delta M_{ij}) ^2 = \frac{N_S-1}{N_S}\sum _S(M_{ij} ^{(s)}-M_{ij})^2,
\label{eq:jk_error}
\end{equation}
where $\mathbfss{M}$ is the covariance matrix $\mathbfss{C}$ or the correlation matrix $\mathbfss{R}$ (for Fig.~\ref{fig:cut_corr} we use $\mathbfss{R}$). $\mathbfss{M} ^{(s)}$ is the covariance or correlation matrix which is obtained when leaving out the $s-$th realization,
\begin{equation}
M_{ij} ^{(s)}= \frac{1}{N_S-1}\sum _{r \neq s}(\xi_i ^{(r)}-\bar{\xi_i})(\xi_j^{(r)}-\bar{\xi_j}).
\end{equation}

For the comparison of the cuts through the correlation matrices, all methods agree well the corresponding N-body measurements with only very small differences.  In order to quantify the discrepancies between the covariance and correlation matrices drawn from the approximate methods to the corresponding N-body measurements, we use an $\chi^2$ approach. Concretely, we compute $\chi^2$ as 
 \begin{equation}
 \chi^2 = \sum _{i} \sum _{j\geq i} \frac{(C_{ij,\text{approx}}-C_{ij,\text{Minerva}})^2}{\Delta C_{ij,\text{Minerva}}^2},
 \label{eq:chi2_cov}
 \end{equation}
 and
  \begin{equation}
 \chi^2 = \sum _{i} \sum _{j> i} \frac{(R_{ij,\text{approx}}-R_{ij,\text{Minerva}})^2}{\Delta R_{ij,\text{Minerva}}^2},
 \label{eq:chi2_corr}
 \end{equation}
 where the indices $i$ and $j$ run over the bins corresponding to the range of interest of $20-160\,h^{-1}$Mpc and $\Delta \mathbfss{ C}_{\text{Minerva}}$ and $\Delta \mathbfss{ R}_{\text{Minerva}}$ are the estimated errors from equation~(\ref{eq:jk_error}).
If the approximate methods perfectly reproduce the expected covariances from the N-body simulations, the $\chi^2$ obtained from the approximate methods should be $\chi^2 \approx 0$ for the predictive and calibrated methods.
This is due to the fact that the simulation boxes of the predictive and calibrated methods match the initial 
conditions of Minerva and therefore the properties of the noise in the estimates of
$\mathbfss{C}$ should be very similar.
For the covariance and correlation matrices obtained from the PDF-based predictions, we expect 
 $\chi^2 \approx N(N-1)/2$ where $N$ is the number of bins of the covariance or correlation matrix, since these predictions do not correspond to the
same initial conditions.
In table \ref{tab:chi2_cov} we list the obtained relative $\chi^2$-values,
\begin{equation}
\chi^2_{rel} = \frac{\chi^2}{N(N-1)/2},
\end{equation}
where $N=42$, for all considered samples and clustering statistics.
We notice that the $\chi^2$-values are in most cases smaller for the predictive than the calibrated methods. Furthermore, the $\chi^2$-values from the wedges measurements are overall smaller than the corresponding ones from the multipole measurements. Also, in most cases the $\chi^2$-values obtained from the covariance matrices are slightly larger than the corresponding ones from the correlation matrices, indicating discrepancies in the variances obtained from the approximated methods.

The computed $\chi^2$-values do not take the covariance between the different entries of $\mathbfss{C}$ into account. In order to provide a more complete picture of how far the multipole and wedges distributions characterized by the different covariance matrices are, we also compute the Kullback-Leibler divergence \citep{Kullback1951,Connell2016}. In our case (two multivariate normal distributions with the same means) the Kullback-Leibler divergence is given as
\begin{equation}
\begin{split}
D_{KL}(\mathbfss{C}_{\text{Minerva}} \parallel \mathbfss{C}_{\text{approx}}) = & \frac{1}{2}
\left(\textrm{tr}(\mathbfss{C}_{\text{approx}}^{-1} \mathbfss{C}_{\text{Minerva}}) \right.\\ &
\left.+\textrm{ln}\left(\frac{\textrm{det} \mathbfss{C}_{\text{approx}}} {\textrm{det} \mathbfss{C}_{\text{Minerva}}}\right) 
-N\right).
\end{split}
\label{eq:dkl_eq}
\end{equation} 
If the approximate methods perfectly reproduce the expected distributions from the N-body simulations, including the same noise, we expect $D_{KL}\approx 0$.
In table \ref{tab:chi2_cov} we list the obtained $D_{KL}$ values. We find that the values for $D_{KL}$ are closer to zero for the predictive than for the other approximate methods. For the calibrated methods and for the distributions with different noise, obtained from the Gaussian and log-normal models, we find values $D_{KL}\approx 1$.

\begin{table*}
	\centering
	\caption{Values of the relative $\chi^2$ for the covariance matrices  $\mathbfss{C}$ (equation \ref{eq:chi2_cov}), correlation matrices $\mathbfss{R}$ (equation \ref{eq:chi2_corr}) and values for the the Kullback-Leibler divergence $D_{KL}$ (equation \ref{eq:dkl_eq}) obtained from the approximate methods.} 
		\label{tab:chi2_cov}
	\begin{tabular}{llllllll}
		\hline
		 code & sample &  $\chi_{rel} ^2$ for $\mathbfss{C}$ from $\xi_{024}$  & $\chi_{rel}  ^2$ for $\mathbfss{C}$ from $\xi_{w}$ & $\chi_{rel} ^2$ for $\mathbfss{R}$ from $\xi_{024}$ &$\chi_{rel}  ^2$ for $\mathbfss{R}$ from $\xi_{w}$ 
& $D_{KL}$ for $\xi_{024}$ & $D_{KL}$ for $\xi_{w}$\\
		\hline
                  \cola & mass1 & 0.19 & 0.21 & 0.17 & 0.16 & 0.24 & 0.24
\\
                  \cola & dens1 & 0.31 &  0.42 & 0.17  & 0.15 & 0.28 & 0.27 \\
                  \cola & bias1 & 0.20 & 0.11 & 0.19 & 0.19 & 0.27& 0.27\\
                  \pin & mass1 & 0.48 & 0.51 & 0.27 & 0.26 & 0.33 & 0.33\\
                  \pin & dens1 & 0.76  & 0.67 & 0.78 & 0.70 & 0.77 & 0.77\\
                   \pin & bias1 & 0.23  & 0.20 & 0.24 & 0.22 & 0.28 & 0.29\\
                  \halogen & mass1 & 1.22 & 0.90 & 1.09 & 0.77 & 1.28 & 1.14\\
                  \patchy & mass1  & 0.67 & 0.40 & 0.73 & 0.44 & 0.82 & 0.79\\
                  Gaussian & mass1  & 2.50 & 2.20 & 2.04 & 0.91 & 0.82 & 1.08\\
                  log-normal & mass1 & 1.76 & 1.09 & 1.31 & 0.97 & 0.96 & 0.98\\
                  & & & & & \\
                  \cola & mass2 & 0.40 & 0.36 & 0.38  & 0.33 & 0.43 & 0.45 \\
                  \cola & dens2 & 0.36 & 0.23 & 0.35  & 0.27 & 0.28 & 0.28\\
                  \pin & mass2 & 1.03 & 1.20 & 0.44 & 0.41 & 0.46 & 0.44\\
                  \pin & dens2 & 0.81 & 0.83 & 0.44 & 0.40 & 0.41 & 0.40\\
                   \pin& bias2 & 0.70 & 0.31 & 0.42 & 0.54 & 0.41 & 0.73\\
                  \peak & mass2 &1.84  &2.02 & 0.69  & 0.69 & 1.05 & 1.03\\
                  \peak & dens2 & 0.48  & 0.47 & 0.48 & 0.45 & 0.46 & 0.48\\
                  \halogen & mass2 &  1.77 &1.32 & 1.70  & 1.29 & 1.07 & 1.07\\
                  \halogen & bias2 & 2.24 &1.76 & 2.06 & 1.59 & 1.28 & 1.32 \\
                  \patchy & mass2 & 1.41  & 1.26 & 1.21  & 0.97 & 0.99 & 1.01\\
                  Gaussian & mass2 & 2.02 & 1.77  & 1.75 & 1.03 & 0.78 & 1.14\\
                  log-normal & mass2 & 2.27  & 2.57 & 1.64 & 1.88 & 1.02 & 1.07 \\
	\end{tabular}
\end{table*}

\subsection{Performance of the covariance matrices}
\label{sec:performance-comp}

For the final validation of the covariance matrices inferred from the different 
approximate methods, we analyse their performance on cosmological parameter constraints. 
We perform fits to the synthetic clustering measurements described in Section~\ref{sec:xi_model}, using the estimates of $\mathbfss{C}$ obtained from the 
different halo samples and approximate methods.
We focus on the constraints on the BAO shift parameters $\alpha_{\parallel}$,  
$\alpha_{\perp}$, and the growth rate $f\sigma_8$. 

\begin{figure*}
\begin{minipage}{\textwidth}
\centering
 \includegraphics[width=0.7\textwidth]{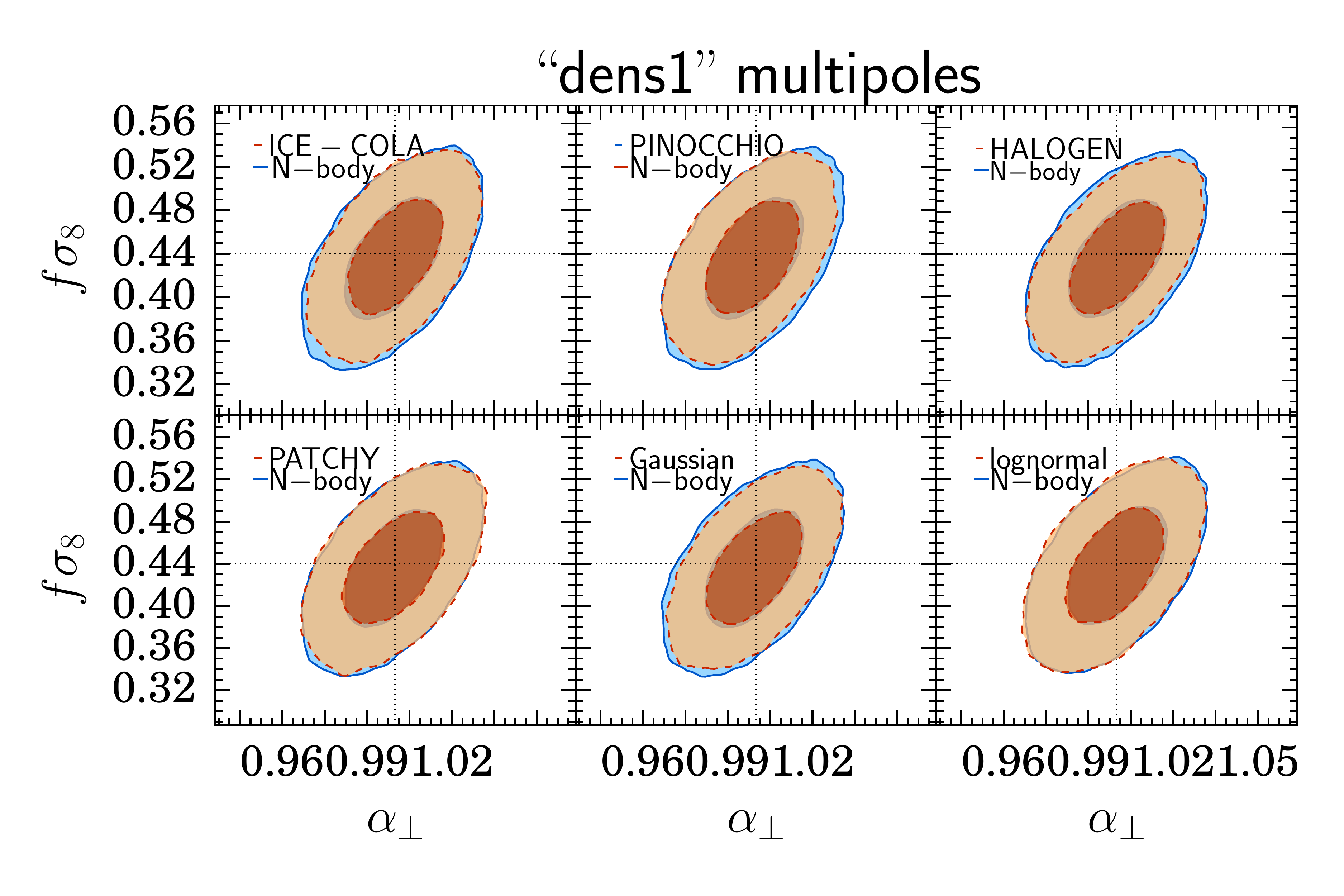}
\end{minipage}
\begin{minipage}{\textwidth}
\centering
 \includegraphics[width=0.9\textwidth]{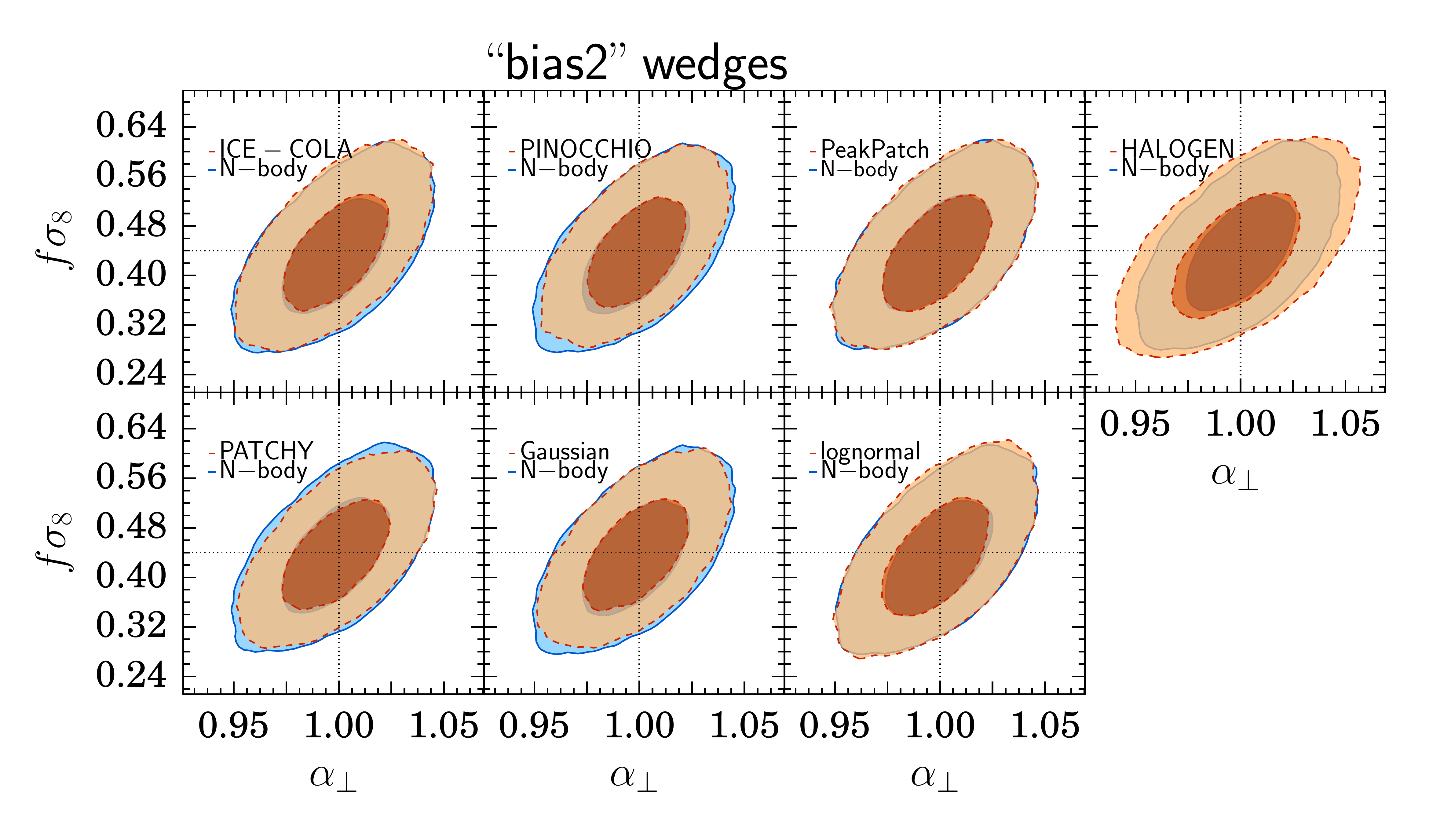}
 \end{minipage}
 \caption{ Comparison of the marginalised two-dimensional constraints in the 
 $\alpha_{\perp}$-$f\sigma_8$ plane for the analysis of samples from the approximate 
 methods with the corresponding constraints obtained from analysis of the parent 
 Minerva sample.
  The contours correspond to the 68\% and 95\% confidence levels.\textit{Upper panel}: 
  Results from the analysis of the multipoles measured from the dens1 samples.
  \textit{Lower panel}: Results from the analysis of the clustering wedges measured from 
  the bias2 samples. }
 \label{fig:constraints_dens1_bias2}
\end{figure*}

Fig.~\ref{fig:constraints_dens1_bias2} shows the 
two-dimensional marginalised constraints in the $\alpha_{\perp}$-$f\sigma_8$ plane 
for the analysis of our two examples cases, the Legendre multipoles measured from 
the dens1 samples (upper panels), and the clustering wedges recovered from the 
bias2 samples (lower panel).

\begin{figure*}
  \begin{minipage}{\textwidth}
  \centering
 \includegraphics[width=0.8\textwidth]{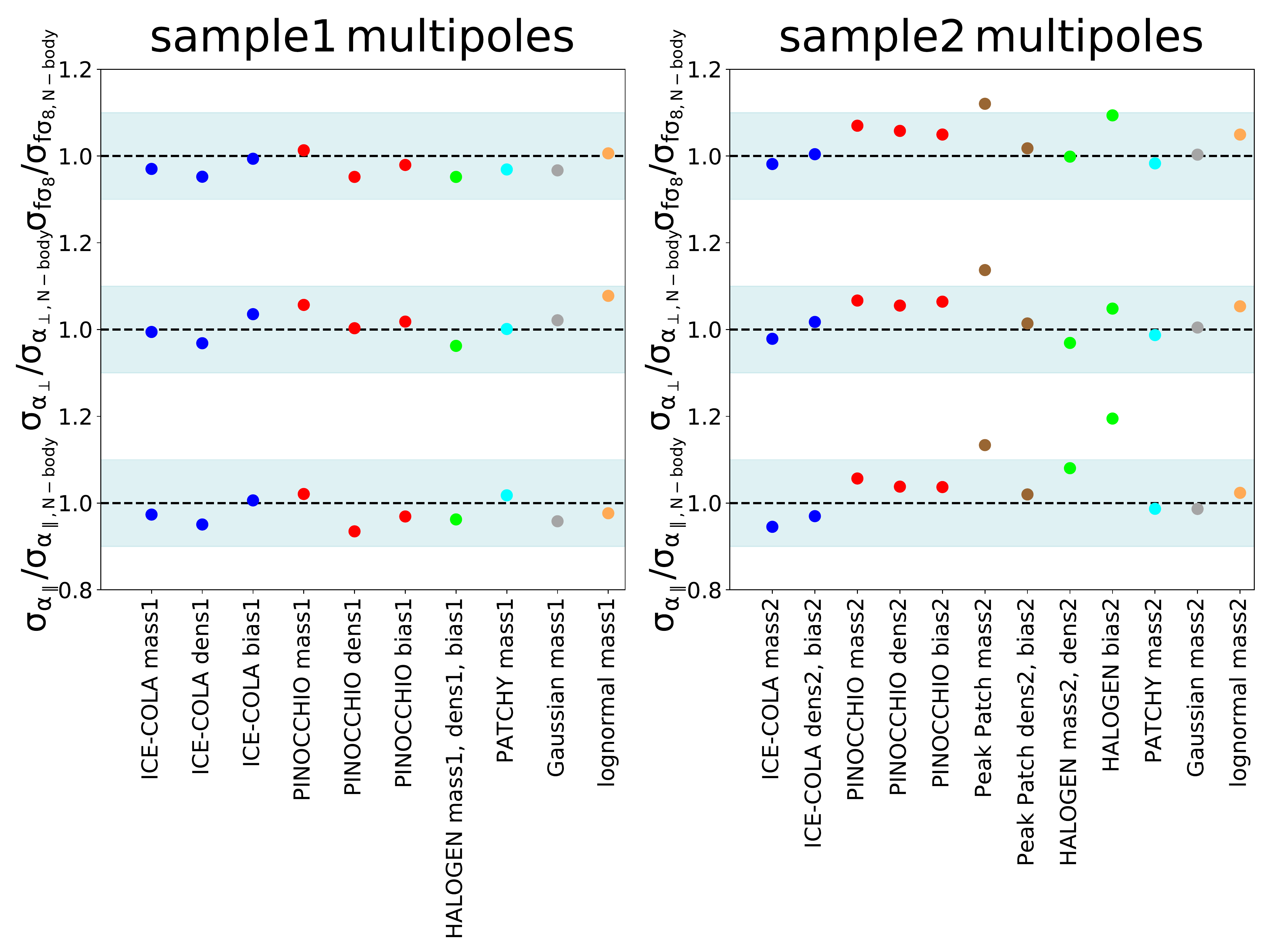}
  \includegraphics[width=0.8\textwidth]{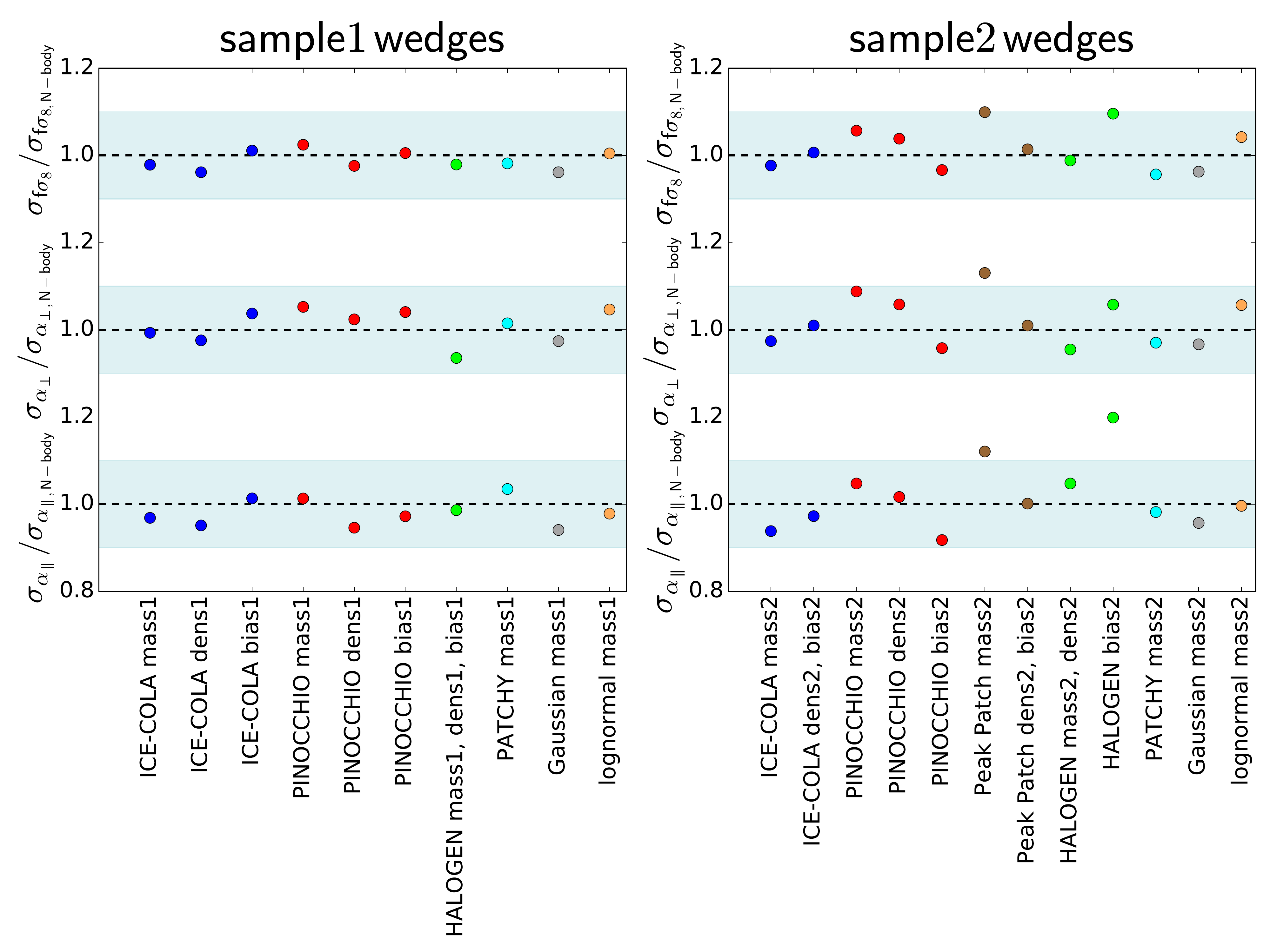}
   \end{minipage}
  \caption{Comparison of the marginalised error on the parameters $\alpha_{\parallel}$, 
  $\alpha_{\perp}$ and $f\sigma_{8}$ which are obtained from the analysis using the 
  covariance matrices from the approximate methods to the corresponding ones from the 
  N-body catalogues. The light grey band indicates a range of $\pm 10\%$  deviation from a 
  ratio equal to 1. The different panels show the results obtained from the analysis of 
  \textit{upper, left panel}: the multipoles drawn from the samples corresponding to 
  the first N-body parent sample with the lower mass cut,  \textit{upper, right panel}:  
  the multipoles drawn from the samples corresponding to the second N-body parent sample 
  with the higher mass cut, \textit{lower, left panel}: the wedges drawn from the samples 
  corresponding to the first N-body sample,  \textit{lower, right panel}: the wedges 
  drawn from the  samples corresponding to the second N-body sample.}
\label{fig:comp_error}
\end{figure*}

In general, the allowed regions for these parameters obtained using the estimates of 
$\mathbfss{C}$ inferred from the different approximate methods (shown by the solid 
lines) agree well with those obtained using the covariance matrices from Minerva
(indicated by the dotted lines in all panels).
However, most cases exhibit small deviations, either slightly under- or over-estimating 
the statistical uncertainties.
We find that, for all samples and clustering statistics, the mean parameter values 
inferred using approximate methods are in excellent agreement with the ones from the 
corresponding N-body analysis, showing differences that are much smaller than their 
associated statistical errors.
The parameter uncertainties recovered using covariances from the approximate methods show 
differences with respect to the N-body constraints ranging between 0.3\% and 
8\% for the low mass samples, while most of the results agree within 5\% with the N-body results,  and between 0.1\% and 20\% for the high-mass cut, while most of the results agree within 10\% with the N-body results.
For the comparison of the obtained parameter uncertainties it is important to point out that in our companion paper \citet{Blot18} estimate that the statistical limit of our parameter estimation is about 4\% to 5\%.
Fig.~\ref{fig:comp_error} shows the ratios of the marginalised parameter errors 
drawn from the analysis with the different approximate methods with respect to the 
N-body results. We observe that for the samples corresponding to the first mass 
cut, all methods reproduce the N-body errors within 10\% for all parameters, and in most cases within 5\% corresponding to the statistical limit of our analysis.
For the samples corresponding to the second mass cut also most methods 
reproduce the N-body errors within 10\% with exception of the \peak mass-matched 
and the \halogen bias-matched samples. 
This might be due to the fact that these two samples have 15-20\% less halos than the corresponding N-body sample.

\begin{figure*}
\begin{minipage}{\textwidth}
\centering
 \includegraphics[width=0.6\textwidth]{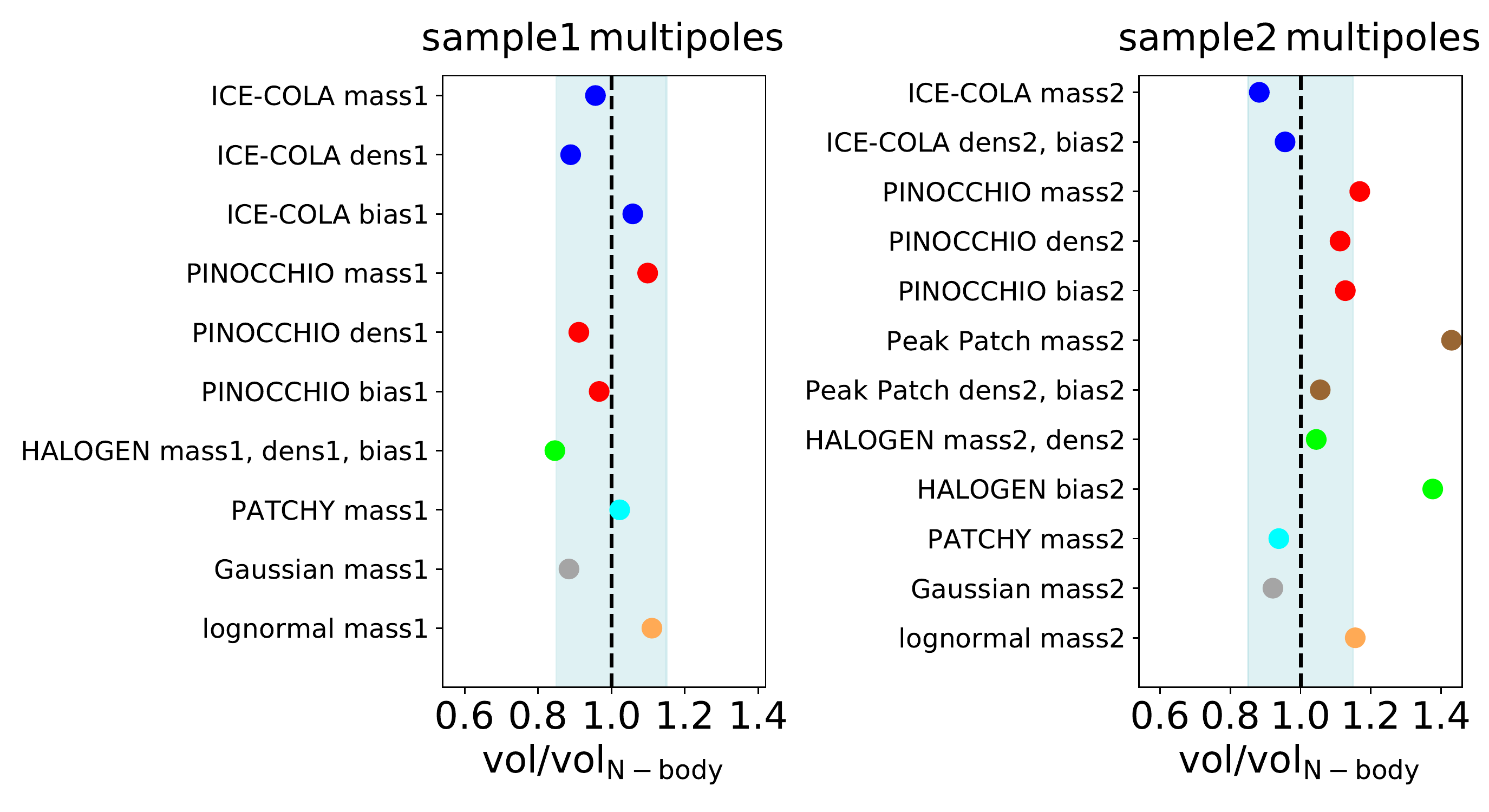}
\end{minipage}
\begin{minipage}{\textwidth}
\centering
  \includegraphics[width=0.6\textwidth]{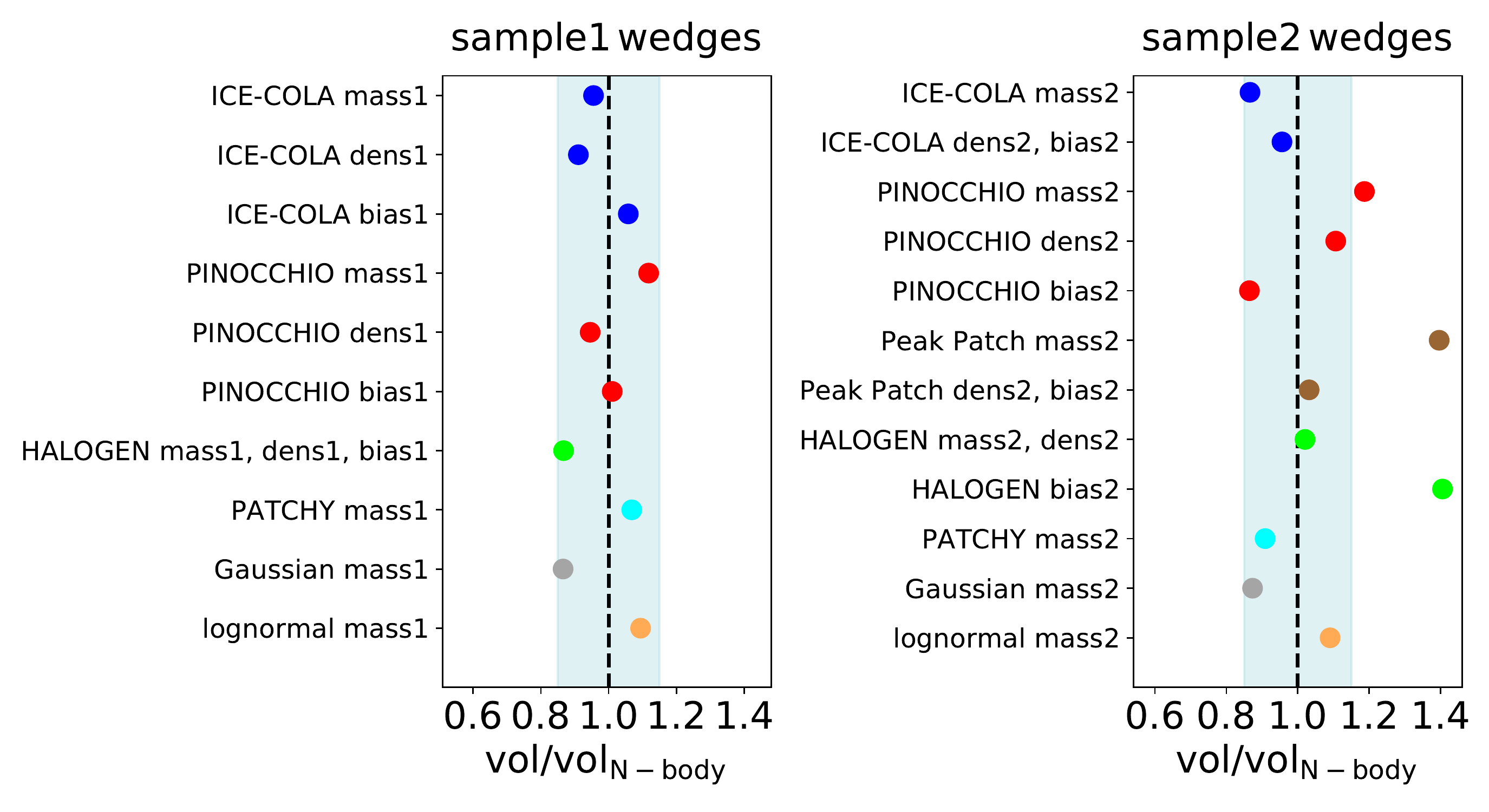}
 \end{minipage}      
   \caption{Comparison of the volume ratios between the allowed statistical volumes 
  obtained from the analysis using the covariance matrices from the approximate methods 
  to the corresponding ones from the N-body catalogues. The light grey band indicates a 
  range of $\pm 10\%$  deviation from a ratio equal to 1. The different panels show the 
  results obtained from the analysis of \textit{upper, left panel}: the multipoles drawn 
  from the samples corresponding to the first N-body parent sample with the lower mass 
  cut,  \textit{upper, right panel}:  the multipoles drawn from the samples corresponding 
  to the second N-body parent sample with the higher mass cut, \textit{lower, left panel}: 
  the wedges drawn from the samples corresponding to the first N-body sample,  
  \textit{lower, right panel}: the wedges drawn from the  samples corresponding to the 
  second N-body sample.}
\label{fig:stat_vol}
\end{figure*}

In order to evaluate the parameter errors, we use the 
volume of the allowed region in the three-dimensional parameter space of 
$\alpha_{\parallel}$, $\alpha_{\perp}$ and $f\sigma_8$, which can be 
estimated as
\begin{equation}
 V = \sqrt{\textrm{det Cov}(\alpha_{\parallel}, \alpha_{\perp}, f\sigma_8)}
\end{equation}
where $\textrm{det Cov}(\alpha_{\parallel}, \alpha_{\perp}, f\sigma_8)$ is the 
determinant of the parameter covariance matrix.
For a Gaussian posterior distribution, the allowed volume is proportional to the volume enclosed 
by the three-dimensional 68\% C.L. contour.
This definition is similar to the two-dimensional Dark Energy Task Force figure of merit 
of the dark-energy equation-of-state parameters $w_0$--$w_a$ 
\citep{Wang2008, Albrecht2006}, but without taking the inverse of the allowed volume.
The ratios of the allowed volumes obtained from the analysis with the different 
approximate methods and the N-body results are shown in Fig.~\ref{fig:stat_vol}. Here the 
differences in the performance of the methods become clearer. For the first mass cut 
we notice that most approximate methods can reproduce the N-body volume at a 10\%~level, 
with the exception of \halogen and the Gaussian and log-normal models, which lead to 
slightly worse results and show 10\%--15\% agreement. For the second mass cut we find overall
larger differences between the samples. The results from the majority of the samples agree 
within 10\% with the N-body results, the rest shows differences of 10\%--15\%, and for the \peak mass2 and \halogen bias2 samples differences of up to 40\%.
For both mass cuts, we find significant differences in the performances of samples 
drawn from the same approximate method but using different selection criteria.

\section{Discussion}
\label{sec:discussion}

In this section we discuss our results on the allowed parameter space volumes  
obtained in Section~\ref{sec:performance-comp}.
Fig.~\ref{fig:stat_vol} clearly shows that there are significant differences in the 
volume ratios between samples drawn from the same approximate method when 
applying different selection criteria to define the halo catalogues. Matching
the parent samples from Minerva by mass limit, number density or 
bias can lead to differences of up to 20\% on the obtained results.

\begin{figure*}
\begin{minipage}{\textwidth}
\centering
 \includegraphics[width=0.7\textwidth]{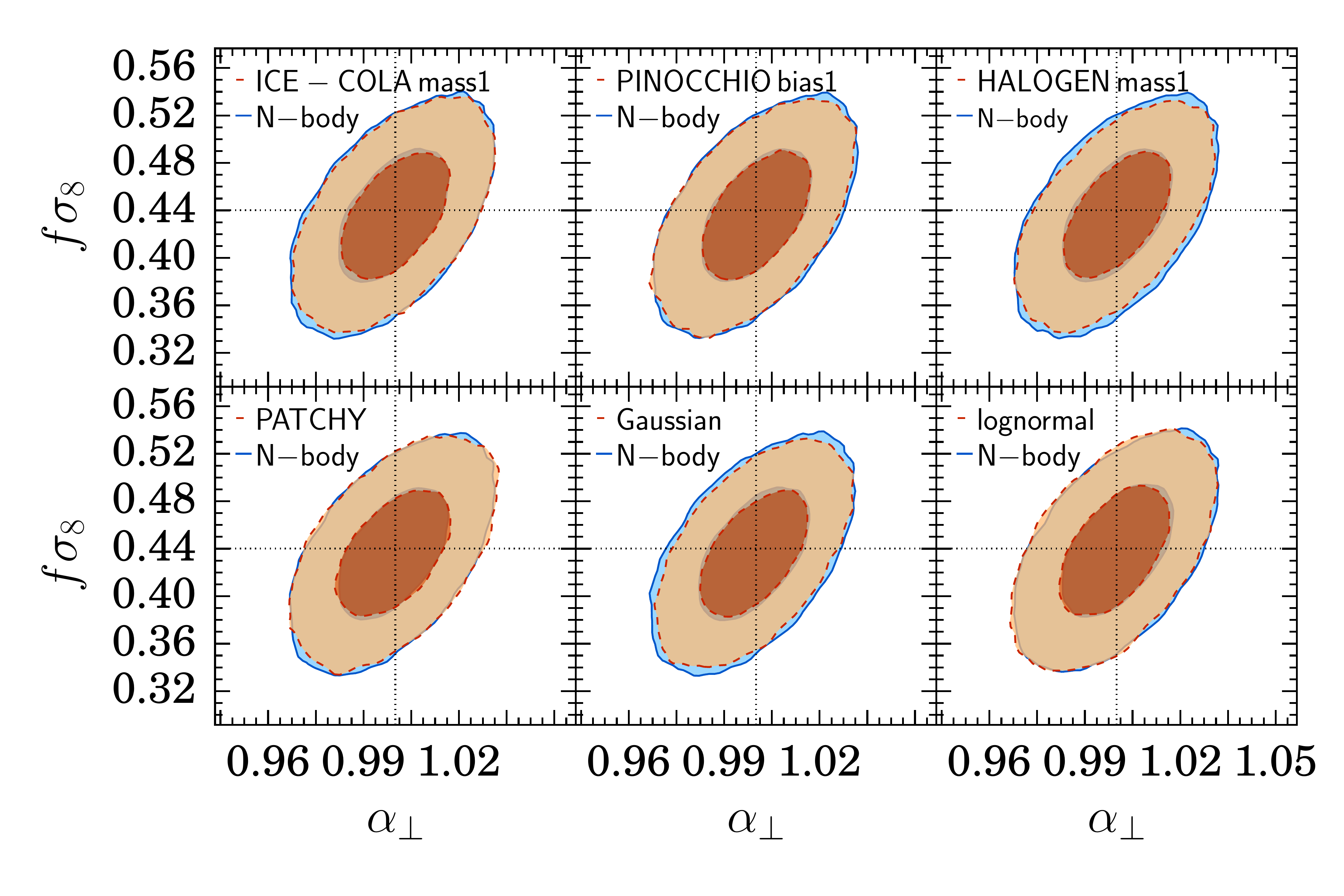}
\end{minipage}
\begin{minipage}{\textwidth}
\centering
 \includegraphics[width=0.8\textwidth]{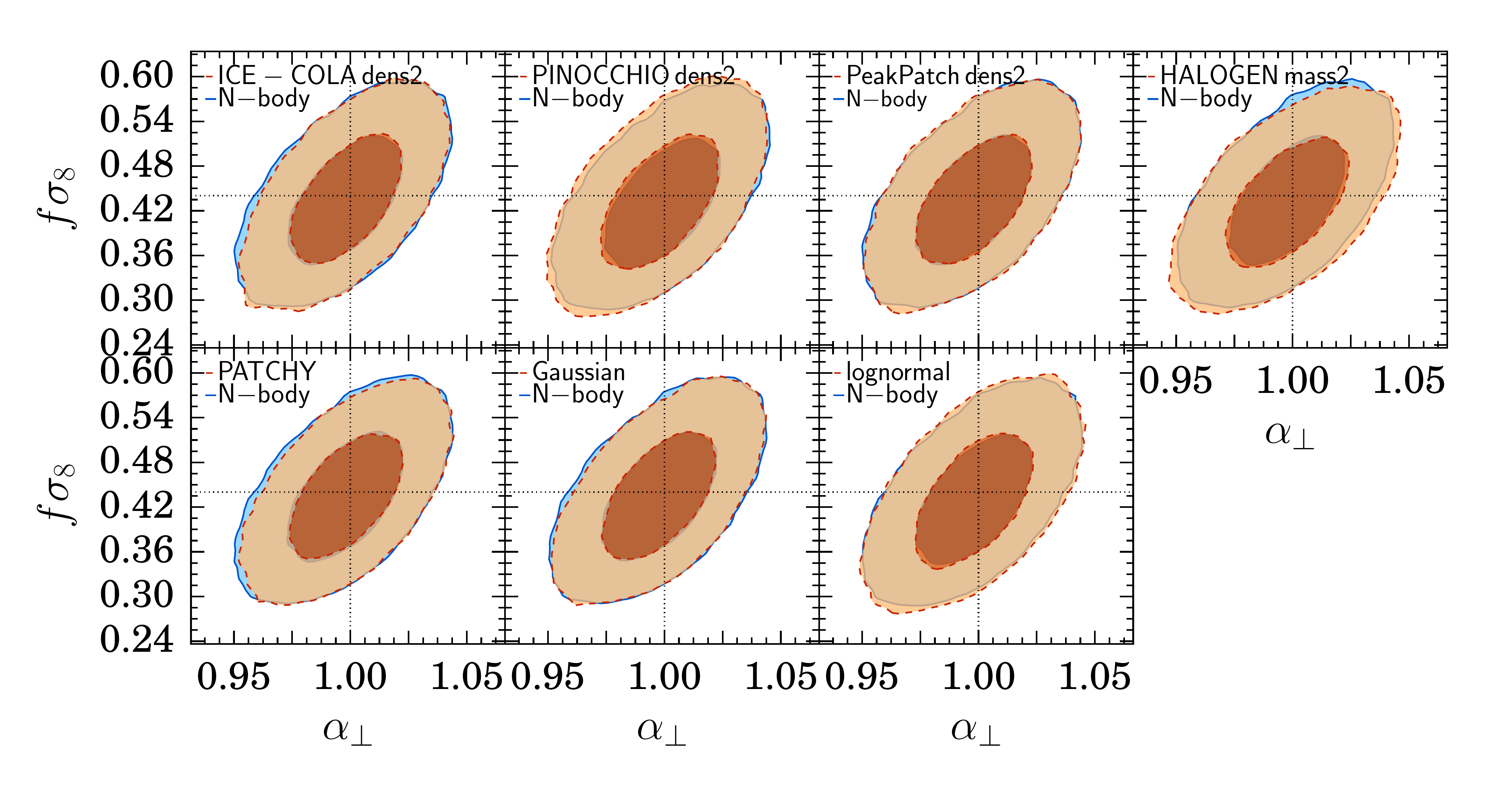}
 \end{minipage}
 \caption{ Comparison of the marginalised two-dimensional constraints in the 
 $\alpha_{\perp}$-$f\sigma_8$ plane for the multipole analysis using the best choice of 
 matching for each approximate method individually to the corresponding constraints 
 obtained from the N-body analysis analysis.
  The contours correspond to the 68\% and 95\% confidence levels.\textit{Upper panel}: 
  Results for the samples corresponding to the first mass cut.
  \textit{Lower panel}:  Results for the samples corresponding to the second mass cut. }
 \label{fig:constraints_best}
\end{figure*}

For each approximate method, mass limit, and clustering statistic, 
we identified the best selection criteria for matching to the N-body parent samples.  
As discussed in Section~\ref{sec:samples}, for \patchy, log-normal and the Gaussian model 
we only have samples characterized by the same mass cuts as the N-body catalogues.
The best cases in decreasing order of the accuracy with which the 
results of the N-body covariances are reproduced are:
\begin{itemize}
\item Lower mass cut, Legendre multipoles:  \patchy ($V/V_{Min} =1.02$), \pin bias matched ($V/V_{Min}  = 0.97$), \cola mass matched ($V/V_{Min}  = 0.96$), log-normal ($V/V_{Min}  = 1.11$), Gaussian ($V/V_{Min} =0.88$), \halogen mass, density, bias matched ($V/V_{Min} =0.85$)
\item Lower mass cut, clustering wedges: \pin bias matched ($V/V_{Min}  =1.01$), \cola mass matched  ($V/V_{Min}  = 0.96$),  \patchy ($V/V_{Min}  = 1.07$), log-normal ($V/V_{Min}  = 1.09$), \halogen mass, density, bias matched ($V/V_{Min}  =0.87$), Gaussian ($V/V_{Min}  =0.87$)
\item Higher mass cut, Legendre multipoles: \cola density matched ($V/V_{Min} =0.96$), \halogen mass, density matched ($V/V_{Min} =1.04$), \peak density, biased matched ($V/V_{Min} =1.06$), \patchy ($V/V_{Min} =0.94$), Gaussian ($V/V_{Min} =0.92$), \pin density matched ($V/V_{Min} =1.11$), log-normal ($V/V_{Min} =1.16$)
\item Higher mass cut, clustering wedges: \halogen mass, density matched ($V/V_{Min} =1.02$), \cola density matched ($V/V_{Min} =0.97$), \peak density, biased matched ($V/V_{Min} =1.03$), \patchy ($V/V_{Min}  =0.91$), log-normal ($V/V_{Min} =1.09$), \pin density matched ($V/V_{Min} =1.1$), Gaussian ($V/V_{Min}  = 0.87$)
\end{itemize}
For a better illustration, Fig.~\ref{fig:constraints_best} shows the two-dimensional
marginalised constraints on $\alpha_{\perp}$ and $f\sigma_8$ obtained from the 
Legendre multipoles for the low (upper panels) and high (lower panels) mass limits.
The different panels show the results obtained from the different approximate methods 
when the best selection criteria for each case is implemented. The overall agreement 
with the results derived from the N-body covariances is better in this case than
when the same definition is applied to all methods.

The best strategy to define the halo samples for a given approximate 
method is often different for our two mass limits.
For example, considering the results from \pin, while for our first mass limit the 
bias-matched halo samples lead to the best agreement with the constraints inferred from 
the N-body covariances, for the second mass threshold the density-matched samples provide 
a better performance.
Focusing on the results from the multipole analysis, we observe that for the first 
mass limit \patchy, \cola and \pin perform slightly better than the 
other methods. These methods reproduce the statistical volume of the allowed parameter 
regions obtained using the N-body covariances within 5\% 
while the other methods only reach a 10\%-15\% agreement.
For the second mass limit \cola, \halogen and \peak can reproduce the N-body results 
within 5\%, \patchy and the Gaussian model within 10\%, and \pin 
and the log-normal model within 15\%. 
It is also interesting to note that the order of performance of the methods 
is slightly different for the multipole and the wedges analysis. For example, the multipole 
analysis using the \patchy covariance matrix leads to a better than 2\% agreement with the 
N-body results, whereas the wedge analysis only reaches 7\%. 

\begin{figure*}
 \includegraphics[width=0.6\textwidth]{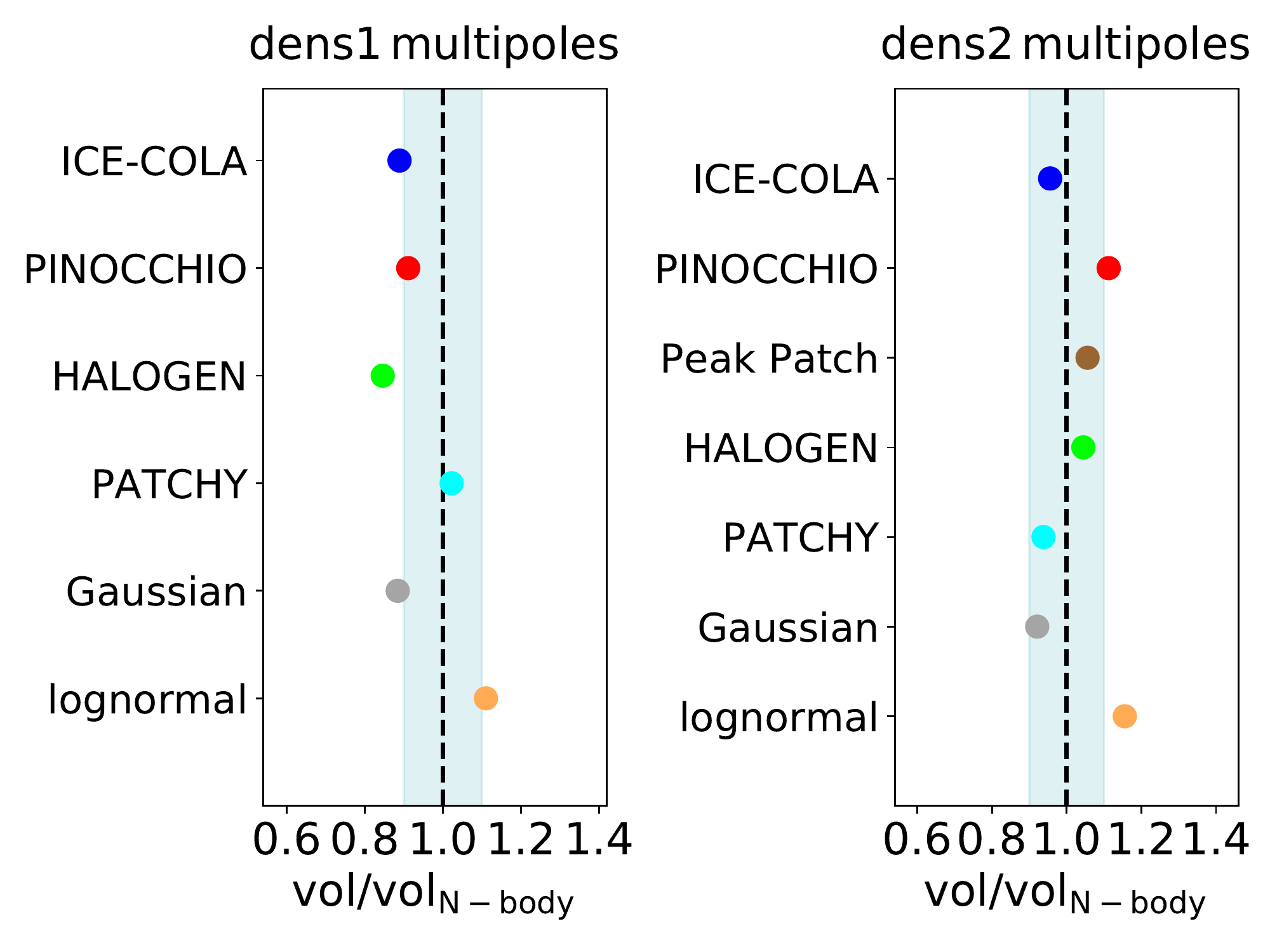}
 \caption{Volume ratios between the allowed statistical volumes obtained from the 
 analysis using the covariance matrices from the approximate methods to the corresponding 
 ones from the N-body catalogues for the density matched samples. The light grey 
 band indicates a range of $\pm 10\%$  deviation from a ratio equal to 1. }
\label{fig:stat_vol_dens}
\end{figure*}

Our analysis is part of a general comparison project of 
approximate methods involving also the covariances of power spectrum and bispectrum 
measurements \citep{Blot18,Colavincenzo18}. 
The power spectrum analysis of \citet{Blot18} is more closely related 
to the one presented here, as it is based on the same baseline model
of the two-dimensional power spectrum and explore constraints on the same nuisance 
and cosmological parameters. The bispectrum covariance analysis of \citet{Colavincenzo18} 
is different in terms of the model and the parameter constraints included in the comparison.
Both of our companion papers consider the same approximate methods and mass cuts used 
here, but focus on the abundance-matched samples.
A comparison of the results of the three studies shows that the differences between 
the predictive, calibrated and PDF-based approximate methods are less evident for 
the correlation function analysis than for the power spectrum and bispectrum. This can 
be clearly seen by comparing the variations of the statistically allowed 
volumes recovered from the different approximate methods when applied to the 
correlation function, power spectrum and bispectrum covariances. 
Since our companion papers focus on the density-matched samples, we also 
show the allowed volumes only for the ``dens'' samples in Fig.~\ref{fig:stat_vol_dens}.
The differences between the approximate methods are less evident in 
configuration space, become more evident for the power spectrum and are strongest 
for the bispectrum analysis. 

In summary, our results and those of our companion papers indicate that 
approximate methods can provide robust covariance matrix estimates for 
cosmological parameter constraints. However, the differences seen between the 
various recipes, statistics, and selection criteria considered here highlight the 
importance of performing detailed tests to find the best strategy to draw halo samples 
from any given approximate method.

\section{Summary and conclusions}
\label{sec:conclusions}
	
We have analysed the performance of several approximate methods at providing 
estimates of the covariance matrices of anisotropic two-point clustering 
measurements in configuration space.
Our analysis is part of a comparison project, including also detailed studies of 
the covariance matrices of power spectrum and bispectrum measurements, 
which are summarized in our companion papers 
\citet{Blot18} and  \citet{Colavincenzo18}, 
respectively.

Our comparison included seven approximate methods, which we divided into three categories: 
predictive methods (\cola, \peak, and \pin), methods that require  
calibration with N-body simulations (\halogen and \patchy), and recipes 
based on assumptions regarding the shape of the density PDF (log-normal and Gaussian 
density fields). We compared these methods against the results obtained 
from the Minerva simulations.
We generated sets of 300 halo catalogues using the predictive and calibrated 
methods, matching the initial conditions of the reference N-body simulations. 
For the log-normal predictions we generated a set of 1000 catalogues designed to 
match the number density and mean correlation function measured 
from the N-body simulations.

We defined two halo samples from the Minerva simulations by applying mass 
thresholds corresponding to 42 and 100 DM particles. 
We then selected different halo samples from the approximate methods by matching the 
mass threshold, number density and clustering amplitude of the parent samples 
from the N-body simulations.
We estimated the covariance matrices of the Legendre multipoles and clustering 
wedges corresponding to all halo samples and compared the results with the 
corresponding ones from the parent catalogues.

Our main comparison was focused on the accuracy with which the covariance matrices
inferred from the approximate methods reproduce the cosmological parameter 
constraints obtained from the N-body results. 
For this, we first used a model of the two-dimensional power spectrum 
applied in recent LSS analyses \citep{Sanchez2017, 
Grieb2017, Salazar-Albornoz2017, Hou2018} to construct synthetic clustering 
measurements, and then fitted these data with the same baseline 
model, using the covariances from the different methods and assuming a Gaussian 
likelihood function. 
We analysed the obtained parameter constraints on 
$\alpha_{\parallel}$, $\alpha_{\perp}$ and $f\sigma_8$.
The mean values obtained from the fits agree perfectly with the N-body results for all 
the samples. Most methods recover the marginalised N-body parameter errors within 5\% for the lower mass cut, which corresponds also to the statistical limit of our analysis,  and 10\% for the higher mass cut.
The comparison of the statistically allowed volumes in the three-dimensional parameter space of 
$\alpha_{\parallel}$, $\alpha_{\perp}$, and $f\sigma_8$ shows that the results obtained 
from any given approximate method by implementing different selection criteria, i.e. by 
matching the mass, number density, or bias of the parent N-body samples, can differ 
by up to 20\%.
Therefore, for each approximate method and mass limit we identified the selection scheme
that provided the closest agreement with the results obtained using the estimates of 
$\mathbfss{C}$  from the N-body simulations. 
For the first mass cut, we found that the methods \cola, \pin and \patchy reproduce the 
N-body results slightly better than the other methods, with differences 
of less than 10\% in the allowed volumes. The remaining methods 
show a 10\%-15\% agreement with the N-body results. For the second mass cut, \cola,
\halogen and \peak perform the best, recovering the N-body allowed volumes within 5\%. The fits using the 
other methods lead to a 5\%-15\% agreement.
It is noteworthy that the simple Gaussian prediction performs similar to the other approximate methods.

We conclude that, with respect to the covariance matrices of configuration-space 
clustering measurements, there is no clear preference for one of the approximate 
methods. The predictive methods \cola, \peak and \pin do not outperform 
the calibrated methods and simpler recipes significantly. The advantage of using the 
calibrated methods is that they are computationally less expensive. However, the 
calibration using full N-body simulations can also be challenging and time-consuming.
In future studies, we will include additional effects, such as the impact of survey 
geometry, that will allow us to extend our analysis to assess the impact 
of applying approximate methods to the analysis of real galaxy surveys.

\section*{Acknowledgements}
This paper and companion papers have benefited of discussions and the stimulating environment of the Euclid Consortium, which is warmly acknowledged.

M. Lippich and A.G. S\'anchez  thank Daniel Farrow, Jiamin Hou and Francesco Montesano for the useful 
discussion. 
M. Lippich and A.G.
S\'anchez acknowledge support from the Transregional Collaborative Research Centre TR33 {\em The Dark Universe} of the German Research Foundation (DFG). 
M. Colavincenzo is supported by the {\em Departments of Excellence 2018 - 2022} Grant awarded by the Italian Ministero dell'Istruzione, dell'Universit\`a e della Ricerca (MIUR) (L. 232/2016), by the research grant {\em The Anisotropic Dark Universe} Number CSTO161409, funded under the program CSP-UNITO {\em Research for the Territory 2016} by Compagnia di Sanpaolo and University of Torino; and the research grant TAsP (Theoretical Astroparticle Physics) funded by the Istituto Nazionale di Fisica Nucleare (INFN). P. Monaco acknowledges support from a FRA2015 grant from MIUR PRIN 2015 {\em Cosmology and Fundamental Physics: illuminating the Dark Universe with Euclid}. P. Monaco and E. Sefusatti acknowledge support from a FRA2015 grant from MIUR PRIN 2015 {\em Cosmology and Fundamental Physics: illuminating the Dark Universe with Euclid} and from Consorzio per la Fisica di Trieste; they are part of the INFN InDark research group. 
L. Blot acknowledges support from the Spanish Ministerio de Econom\'ia y Competitividad (MINECO) grant ESP2015-66861. M.Crocce acknowledges support from the Spanish Ram\'on y Cajal MICINN program. M.Crocce has been funded by AYA2015-71825. 

C. Dalla Vecchia acknowledges support from the MINECO through grants AYA2013-46886, AYA2014-58308 and RYC-2015-18078. S. Avila acknowledges support from the UK Space Agency through grant ST/K00283X/1. A. Balaguera-Antol\'{i}nez acknowledges financial support from MINECO under the Severo Ochoa program SEV-2015-0548. M. Pellejero-Ibanez acknowledges support from MINECO under the grand AYA2012-39702-C02-01. P. Fosalba acknowledges support from MINECO through grant ESP2015-66861-C3-1-R and Generalitat de Catalunya through grant 2017-SGR-885. A. Izard was supported in part by Jet Propulsion Laboratory, California Institute of Technology, under a contract with the National Aeronautics and Space Administration. He was also supported in part by NASA ROSES 13-ATP13-0019, NASA ROSES 14-MIRO-PROs-0064, NASA ROSES 12- EUCLID12-0004, and acknowledges support from the JAE program grant from the Spanish National Science Council (CSIC). R. Bond, S. Codis and G. Stein are supported by the Canadian Natural Sciences and Engineering Research Council (NSERC). G. Yepes acknowledges financial support from MINECO/FEDER (Spain) under research grant AYA2015-63810-P. 

The Minerva simulations have been performed and analysed on the Hydra and Euclid clusters at the Max Planck Computing and Data Facility (MPCDF) in Garching.

\pin mocks were run on the GALILEO cluster at CINECA thanks to an agreement with  the University of Trieste.

\cola simulations were run at the MareNostrum supercomputer - Barcelona Supercomputing Center (BSC-CNS, www.bsc.es), through the grant AECT-2016- 3-0015. 

\peak simulations were performed on the GPC supercomputer at the SciNet HPC Consortium. SciNet is funded by: the Canada Foundation for Innovation under the auspices of Compute Canada; the Government of Ontario; Ontario Research Fund - Research Excellence; and the University of Toronto.

Numerical computations with \halogen were done on the Sciama High Performance Compute (HPC) cluster which is supported by the ICG, SEPNet and the University of Portsmouth.

\patchy mocks have been computed  in part at the MareNostrum supercomputer of the Barcelona Supercomputing Center thanks to a grant from the Red Espa\~nola de Supercomputaci\'on (RES), and in part at the Teide High-Performance Computing facilities provided by the Instituto Tecnol\'ogico y de Energ\'{\i}as Renovables (ITER, S.A.).




\bibliographystyle{mnras}
\bibliography{reference}



\appendix


\bsp	
\label{lastpage}
\end{document}